\documentclass[aps,showpacs,twocolumn,prd,superscriptaddress,nofootinbib]{revtex4-1}

\usepackage{amsmath}
\usepackage{amsfonts}
\usepackage{amssymb}
\usepackage{latexsym}
\usepackage{graphicx}
\usepackage{bm}
\usepackage[usenames, dvipsnames]{color}
\usepackage{enumerate}
\usepackage[normalem]{ulem} %I like \emph to italicize text instead of underline, hence the normalem option
\usepackage{array}

\newcommand{\be}{\begin{equation}}
\newcommand{\ee}{\end{equation}}
\newcommand\ud{{\mathrm{d}}}

\newcommand\calO{{\mathcal{O}}}
\newcommand\calS{{\mathcal{S}}}
\newcommand\calF{{\mathcal{F}}}
\newcommand\calA{{\mathcal{A}}}
\newcommand\calC{{\mathcal{C}}}
\newcommand\calN{{\mathcal{N}}}

\newcommand{\ph}[1]{\phantom{#1}}

\newcommand{\nn}{\nonumber}

\newcommand{\Msol}{M_{\odot}}

\renewcommand{\v}[1]{\ensuremath{\mathbf{#1}}} % for vectors
\newcommand{\gv}[1]{\ensuremath{\mbox{\boldmath$ #1 $}}} % for vectors with greek symbols
\newcommand{\uv}[1]{\ensuremath{\mathbf{\hat{#1}}}} % for unit vector

\newcolumntype{C}[1]{>{\centering\arraybackslash}p{#1}}
\newcolumntype{L}[1]{>{\raggedright\arraybackslash}p{#1}}

\begin{document}

\title{Gravitational waveforms in scalar-tensor gravity at 2PN relative order}

\author{Noah Sennett}
\email{nsennett@umd.edu}
\affiliation{Department of Physics, University of Maryland, College Park, Maryland 20742, USA}
\affiliation{Max Planck Institute for Gravitational Physics (Albert Einstein Institute), Am M\"uhlenberg 1, Potsdam-Golm, 14476, Germany}
\author{Sylvain Marsat}
\affiliation{Max Planck Institute for Gravitational Physics (Albert Einstein Institute), Am M\"uhlenberg 1, Potsdam-Golm, 14476, Germany}
\affiliation{Department of Physics, University of Maryland, College Park, Maryland 20742, USA}
\affiliation{Gravitational Astrophysics Laboratory, NASA Goddard Space Flight Center, Greenbelt, MD 20771}
\author{Alessandra Buonanno}
\affiliation{Max Planck Institute for Gravitational Physics (Albert Einstein Institute), Am M\"uhlenberg 1, Potsdam-Golm, 14476, Germany}
\affiliation{Department of Physics, University of Maryland, College Park, Maryland 20742, USA}

\date{\today}

\begin{abstract}

We compute the gravitational waveform from a binary system in scalar-tensor gravity at 2PN relative order. We restrict our calculation to non-spinning binary systems on quasi-circular orbits and compute the spin-weighted spherical modes of the radiation. The evolution of the phase of the waveform is computed in the time and frequency domains. The emission of dipolar radiation is the lowest-order dissipative process in scalar-tensor gravity. However, stringent constraints set by current astrophysical observations indicate that this effect is subdominant to quadrupolar radiation for most prospective gravitational-wave sources.  We compute the waveform for systems whose inspiral is driven by: (a) dipolar radiation (e.g., binary pulsars or spontaneously scalarized systems) and (b) quadrupolar radiation (e.g., typical sources for space-based and ground-based detectors).

\end{abstract}

%TO BE ADAPTED

\maketitle

%%%%%%%%%%%%%%%%%%%%%%%%%%%%%%%%%%%%

\section{Introduction}
\label{sec:intro}
The observation of gravitational-wave (GW) events GW150914 and GW151226 by Advanced LIGO marks the dawn of GW astronomy~\cite{Abbott2016a,Abbott:2016nmj}. We expect to observe several such events per year~\cite{Aasi2013,Abbott2016b,DelPozzo2011,TheLIGOScientific:2016pea} with the upcoming network of ground-based detectors comprised of Advanced LIGO~\cite{Abbott2016c}, Advanced VIRGO~\cite{Accadia2010}, KAGRA~\cite{Yoichi2013}, and LIGO-India~\cite{Iyer2011}. These ground-based detectors can observe binary systems containing neutron stars and/or stellar-mass black holes (with a total mass $M{\sim1-100}\Msol$); future space-based detectors like the proposed eLISA mission~\cite{AmaroSeoane2013} will observe binary systems composed of intermediate-mass and/or supermassive black holes ($M\sim100-10^7 \Msol$). Gravitational-wave observations allow us to not only measure the astrophysical properties of these systems but can also be used to test general relativity (GR). Because the coalescence of a compact binary system produces extreme gravitational fields that vary over short time scales, observations of such events allow us to probe the highly-dynamical, strong-field regime of gravity for the first time~\cite{Abbott2016d,TheLIGOScientific:2016pea}.

The detection and analysis of GWs with ground-based detectors require banks of very accurate template waveforms. The prospects of testing gravity with these detectors hinge on our ability to model waveforms in both GR and alternative theories of gravity. Given a GW detection, one can adopt either a \emph{theory-independent} or \emph{theory-dependent} approach to testing GR. A theory-independent test employs waveforms that deviate from a GR signal in some generic, parameterized manner (for examples, see Refs.~\cite{Arun2006b,Yunes2009,DelPozzo2011}). One compares an observed GW signal against these template waveforms to constrain the deviations from GR. A theory-dependent test instead uses waveforms predicted in a particular alternative theory of gravity, comparing them against the detected GW to estimate the underlying physical parameters of that theory. Each approach has its advantages: theory-independent tests can constrain a wide range of alternative theories while theory-dependent tests can directly constrain the fundamental physics of an alternative theory. Both types of tests were performed for GW150914 and GW151226 by the LIGO and Virgo collaborations in Refs.~\cite{Abbott2016d,TheLIGOScientific:2016pea}. For a comprehensive review of proposed theory-independent and theory-dependent tests, see Refs.~\cite{Will2014,YunesLR} and references therein.

In this paper, we present waveforms in scalar-tensor theories of gravity suitable for theory-dependent tests of GR. In particular, we construct ready-to-use waveforms for the inspiral of non-spinning binary systems accurate up to second post-Newtonian (2PN) order, i.e., $\calO\left((v/c)^4\right)$ beyond leading order.\footnote{We describe post-Newtonian (PN) corrections of order $\calO(c^{-2n})$ as ``nPN,'' which we also abbreviate with the notation $\calO(2n)$.} We restrict our attention to systems on quasi-circular orbits, as binaries formed in the field are expected to radiate away any initial eccentricity at frequencies too low to be observable by GW detectors.

Scalar-tensor theories are amongst the most natural alternatives to GR. Specifically, we focus on theories where a single massless scalar $\phi$ non-minimally couples to the metric $g_{\mu \nu}$. Written in the Jordan frame, the action for such theories is given by
\begin{align}
S&=\int d^4 x \frac{\sqrt{-g}}{2 \kappa}\left[\phi R-\frac{\omega(\phi)}{\phi} g^{\mu \nu} \nabla_\mu \phi \nabla_\nu \phi\right]+S_m[g_{\mu \nu},\Xi],\label{eq:JFaction}
\end{align}
where $\kappa=8 \pi G_{*}/c^3$ depends on the bare gravitational coupling constant $G_{*}$. The action for the matter in the theory $S_m$ is a function of only the metric and matter degrees of freedom $\Xi$; the scalar field does not couple to matter directly, only indirectly through its interactions with the metric.
%\footnote{We work in units where $G_{*}=c=1$, but occasionally restore powers of $c$ for the sake of clarity.} 

The restricted class of scalar-tensor theories described by Eq.~\eqref{eq:JFaction} has been studied extensively in the literature because it is general enough to manifest many different deviations from GR yet simple enough that its predictions can be worked out completely.  At 2PN order, the Fokker Lagrangian for a system of point particles was first computed in Ref.~\cite{Damour1996} using an effective field-theory approach. The 2PN metric and equations of motion were computed for bodies composed of perfect fluids in Ref.~\cite{Xie2007}. The post-Minkowskian technique of direct integration of the relaxed Einstein equations (DIRE) was used in a recent series of papers \cite{Mirshekari2013,Lang2014a,Lang2015a} to compute the equations of motion for a system of compact objects at 2.5PN order, as well as the gravitational waveform and energy flux at 2PN (relative) order for binaries on generic orbits. For comparison, the entire waveform for non-spinning systems in GR is known at 3PN order~\cite{BFIS08}, and its quadrupolar and octupolar parts are known at 3.5PN order~\cite{FMBI12, FBI14} (see Ref.~\cite{Bliving} for a review of existing results in GR).

In this paper, we specialize the results of Refs.~\cite{Mirshekari2013,Lang2014a,Lang2015a} to binary systems on quasi-circular orbits and present the waveform in a form that can be easily used to test GR with GWs. This calculation serves as an extension of Ref.~\cite{Will1994}, in which the leading-order behavior of the GW signal produced by binary systems was computed in Brans-Dicke theory \cite{Jordan,Fierz,Brans}, where the scalar coupling $\omega(\phi)=\omega_\text{BD}$ is constant. This work extends those earlier findings to higher PN order in a larger class of scalar-tensor theories.

The paper is organized as follows. In Sec.~\ref{sec:formalism}, we review some preliminary information regarding the production and detection of GWs in scalar-tensor theories. Section~\ref{sec:dynamics} presents the dynamics for binary systems on quasi-circular orbits. In Sec.~\ref{sec:hered}, we compute the hereditary contributions to the gravitational waveform from such systems. We present the binding energy and energy flux in Sec.~\ref{sec:flux} and compute the associated orbital phase evolution. In Sec.~\ref{sec:modes}, we decompose the waveform into spin-weighted spherical modes, and in Sec.~\ref{sec:spa}, we express these modes in Fourier space using the stationary phase approximation. We provide some concluding remarks in Sec.~\ref{sec:conclusions}. Appendix~\ref{app:notation} details the conversion of our notation to that of Refs. \cite{Damour1992,Damour1996}, which is commonly found in the literature. Appendix~\ref{app:coeffs} contains formulae omitted from the main text for the sake of compactness.

All calculations are done both for systems whose inspiral is driven by dipolar radiation and for those driven by quadrupolar radiation; this distinction is discussed in detail in Sec.~\ref{sec:constraints}. Note that the \emph{complete} 2PN (relative) order results are given for only the former case (dipolar-radiation driven systems). The results for the latter case depend on higher-order corrections to the energy flux that have not yet been computed, but we argue in Sec.~\ref{subsec:phasingQD} that the impact of these missing terms is very small.

Henceforth, we work in units where $G_{*}=c=1$.
%%%%%%%%%%%%%%%%%%%%%%%%%%%%%%%%%%%%

\section{Gravitational waves in scalar-tensor gravity}
\label{sec:formalism}
This section contains information concerning the generation of GWs in scalar-tensor gravity that will prove useful throughout the rest of the paper despite not directly contributing to the computation of the waveform. We begin by discussing the behavior of binary systems, tracing new phenomena not found in GR to violations of the strong equivalence principle. We then review the current experimental constraints set on these theories and show that, in most cases, sources for ground- and space-based GW detectors evolve similarly to as in GR. Finally, we discuss the response of a detector to a GW in scalar-tensor gravity and delineate the waveform computed in the subsequent sections.

\subsection{Binary systems of compact objects}
Differences between the dynamics and GW emission of binary systems in scalar-tensor gravity and those in GR ultimately stem from the non-minimal coupling between the metric and scalar field. As a result, the gravitational ``constant'' experienced by massive bodies depends on the value of the background scalar field in which they are situated. For test bodies, this dependence can be deduced directly from Eq.~\eqref{eq:JFaction}: the strength of their gravitational interaction scales as $\phi^{-1}$.

\begin{table*}[t]
\begin{ruledtabular}\caption{Parameters that govern gravitational wave production in binary systems. Quantities listed with the subscript $0$ are evaluated at the value of the background scalar field $\phi_0$.}\label{table:parameters}
\begin{tabular}{L{.1\textwidth}C{.24\textwidth}L{.15\textwidth}L{.51\textwidth}}
Parameter&Defintion&Parameter&Definition\\
\hline
\multicolumn{2}{l}{\emph{Weak-field parameters}}&\multicolumn{2}{l}{\emph{Binary parameters}}\\
%$G$ & $\phi_0^{-1}(4 + 2 \omega_0)/(3+2\omega_0)$ & &\\
%$\zeta$ & $1/(4+2\omega_0)$ & &\\
%$\lambda_1$ & $(d \omega/d\varphi)_0\zeta^2/(1-\zeta)$ & &\\
%$\lambda_2$ & $(d^2 \omega/d\varphi^2)_0\zeta^3/(1-\zeta)$ & &\\
$G$&$G_{*}\phi_0^{-1} (4+2\omega_0)/(3+2\omega_0)$&\multicolumn{2}{l}{\emph{Newtonian}}\\
$\zeta$&$1/(4+2\omega_0)$&$\qquad\alpha $&$1 - \zeta + \zeta (1-2s_1)(1- 2s_2) $
\\
$\lambda_1$&$(\ud\omega/\ud\phi)_0 \phi_0 \zeta^2/(1-\zeta)$&\multicolumn{2}{l}{\emph{post-Newtonian}}\\
$\lambda_2$&$(\ud^2\omega/\ud\phi^2)_0 \phi_0^2\zeta^3/(1-\zeta)$&$\qquad\gamma$ & $-2 \alpha^{-1}\zeta (1-2s_1)(1-2s_2)$
\\
&&$\qquad\beta_1 $&$\alpha^{-2} \zeta (1-2s_2)^2 \left ( \lambda_1 (1-2s_1) + 2 \zeta s'_1 \right )$
\\
\multicolumn{2}{l}{\emph{Strong-field parameters}}&$\qquad\beta_2 $&$\alpha^{-2} \zeta (1-2s_1)^2 \left ( \lambda_1 (1-2s_2) + 2 \zeta s'_2 \right )$
\\
$s_A$&$[\ud \ln m_A(\phi)/\ud \ln \phi]_0$&\multicolumn{2}{l}{\emph{2nd post-Newtonian}}\\
$s'_A$&$[\ud^2 \ln m_A(\phi)/\ud \ln \phi^2]_0$&$\qquad\delta_1$ &$ \alpha^{-2} \zeta (1-\zeta) (1-2s_1)^2$
\\
$s''_A$&$[\ud^3 \ln m_A(\phi)/\ud \ln \phi^3]_0$&$\qquad\delta_2$ &$\alpha^{-2} \zeta (1-\zeta) (1-2s_2)^2 $
\\
 &&$\qquad\chi_1 $&$ \alpha^{-3} \zeta (1-2s_2)^3  \left [ (\lambda_2 -4\lambda_1^2 + \zeta \lambda_1 ) (1-2s_1) -6 \zeta \lambda_1 s'_1 + 2 \zeta^2 s''_1 \right ]
$
\\
&& $\qquad\chi_2 $&$ \alpha^{-3} \zeta (1-2s_1)^3  \left [ (\lambda_2 -4\lambda_1^2 + \zeta \lambda_1 ) (1-2s_2) -6 \zeta \lambda_1 s'_2 + 2 \zeta^2 s''_2 \right ] $
\end{tabular}
\end{ruledtabular}
\end{table*}

The gravitational interaction between compact, self-gravitating bodies is more complex. Because the binding energy of a single self-gravitating body depends on the interactions between all of its constituents, the body's mass $m_A(\phi)$ depends on the local scalar field. This phenomenon is a manifestation of the violation of the strong equivalence principle, as the self-interaction of a massive body is dictated by its composition. As is done in the literature, we adopt an approach proposed by Eardley~\cite{Eardley1975} to handle the interplay between microphysics and gravity that determines the connection between the body's composition and $m_A(\phi)$. We treat compact objects as point particles whose mass is given by $m_A(\phi)$. Rather than solve for this function outright, we parameterize it by its expansion about a background field $\phi_0$
\begin{align}
m_A(\phi)&=m_A^{(0)}\left[1+s_A \Psi+\frac{1}{2}\left(s_A^2+s'_A-s_A\right)\Psi^2+\cdots\right],
\end{align}
where we've defined
\begin{align}
m_A^{(0)}\equiv&m_A(\phi_0),\\
s_A\equiv&\left(\frac{\ud \ln m_A}{\ud \ln \phi}\right)_{\phi=\phi_0},\\
s'_A\equiv&\left(\frac{\ud^2 \ln m_A}{\ud (\ln \phi)^2}\right)_{\phi=\phi_0},\\
\Psi\equiv&\frac{\phi-\phi_0}{\phi_0}.
\end{align}
The parameter $s_A$ is known as the sensitivity of the body. For test bodies, $s_A=0$, while for stationary black holes, $s_A=1/2$~\cite{Hawking1972}.\footnote{The sensitivity of neutron stars is often estimated to be of the order $\sim 0.2$. While true in Brans-Dicke theory~\cite{Eardley1975,Will1989} and some slight variations~\cite{Zaglauer1992}, this result does not hold for generic choices of $\omega(\phi)$. One of the most popular classes of scalar-tensor theories, those that allow spontaneous~\cite{Esposito-Farese1993} and dynamical scalarization~\cite{Barausse2013,Shibata2014}, are a striking counterexample. In these theories, neutron-star sensitivities can be large and \emph{negative}; the process of spontaneous scalarization describes stars whose sensitivity diverges, i.e., $s_A \rightarrow -\infty$.}

The underlying parameters that govern the orbital dynamics and gravitational emission of binary systems up to 2PN order are given on the left-hand side of Table~\ref{table:parameters}. These parameters are classified as either \emph{weak-field} or \emph{strong-field}: the former class influence behavior in all gravitational contexts whereas the latter class only enter in systems with strong gravitational fields, such as those found in self-gravitating compact objects. The weak- and strong-field parameters appear in only a small set of combinations, denoted as the \emph{binary parameters} in Table~\ref{table:parameters}. We have adopted the notation introduced in Refs.~\cite{Mirshekari2013}; the mapping between these parameters and the notation used in Refs. \cite{Damour1992,Damour1996} is given in Appendix \ref{app:notation}.

Novel behavior in scalar-tensor gravity stems from violations of the strong equivalence principle, and thus, is dictated by the strong-field parameters. For example, dipolar emission, the most prominent new effect not found in GR, is tied to $(s_1-s_2)^2$. Formally, dipolar radiation is generated at one PN order lower than quadrupolar radiation (the dominant dissipative channel in GR); in keeping with the conventions of Refs. \cite{Mirshekari2013,Lang2014a,Lang2015a}, we demarcate dipolar emission as a $-1$PN order effect.

\subsection{Generic constraints on scalar-tensor gravity}\label{sec:constraints}

A hundred years of tests have confirmed that gravity closely resembles GR~\cite{Will2014}. Restricting our attention to only those theories that satisfy these constraints, we must study the regime in which new scalar-tensor effects are small relative to those also found in GR. In this limit, the structure of the PN expansion is modified; for example, in the frequency band of interest, the dominant dissipative process is the emission of Newtonian order quadrupolar radiation rather than the $-$1PN dipolar energy flux. We investigate which systems fall within this regime by first mapping the current constraints on scalar-tensor theories to the parameters given in Table~\ref{table:parameters}.

%Thus, the most relevant application of our results are not to \emph{generic} scalar-tensor theories, but instead only to those consistent with these tight experimental constraints. Expressed equivalently, we wish to study the waveforms constructed above in the \emph{near-GR regime}, in which the impact of new scalar-tensor effects are small relative to those also found in GR. The structure of the waveform is fundamentally different in this regime; for example, the dominant dissipative process is the emission of Newtonian order quadrupolar flux rather than the $-$1PN dipolar flux.

%The underlying parameters that govern the GW signal produced by binary systems up to 2PN are given on the left-hand side of Table~\ref{table:parameters}. These parameters are classified as either \emph{weak-field} or \emph{strong-field}: the former class influence behavior in all gravitational contexts whereas the latter class only enter in systems with strong gravitational fields, such as those found in self-gravitating compact objects. The best constraints on both classes of parameters come from a combination of solar-system experiments and binary-pulsar observations.

The best constraints on weak- and strong-field parameters come from a combination of solar-system experiments and binary-pulsar observations. The weak-field parameters $G,\zeta,\lambda_1,\lambda_2$ are tied to the behavior of the scalar coupling $\omega(\phi)$ near the background value of the scalar field $\phi_0$. These quantities can be expressed in terms of the parametrized post-Newtonian (PPN) parameters $\gamma_\text{PPN}$ and $\beta_\text{PPN}$ as well as the 2PN parameter $\epsilon$ introduced in Ref.~\cite{Damour1996}
\begin{align}
G=&\frac{2}{(1+\gamma_\text{PPN})\phi_0} ,\label{eq:GPPN}\\
\zeta=&\frac{1-\gamma_\text{PPN}}{2} ,\\
\bar\lambda_1=& \frac{2\sqrt{2}(\beta_\text{PPN}-1)\phi_0}{\sqrt{1+\gamma_\text{PPN}}},\\
\bar\lambda_2=&\frac{(\epsilon(\gamma_\text{PPN}-1)+24(\beta_\text{PPN}-1)^2 )\phi_0^2}{1+\gamma_\text{PPN}}\label{eq:lambda2PPN},
\end{align}
where we have used the rescaled parameters $\bar\lambda_1\equiv \lambda_1 \sqrt{\zeta}$ and $\bar\lambda_2\equiv \lambda_2 \zeta$ because $\lambda_1$, $\lambda_2$ are not well defined in the GR limit.

The current constraints on these parameters are given in Table \ref{table:constraints}. The constant background field $\phi_0$ is undetectable with weak-field measurements --- at Newtonian order, a redefinition of the field $\phi\rightarrow \phi/\phi_0$ can be compensated by the rescaling of the bare gravitational constant $G_*\rightarrow G_*/\phi_0$ and the redefinition $\omega\rightarrow \phi_0 \omega$. For simplicity, we set $\phi_0$ to unity in Table \ref{table:constraints}. Note that the constraint on $\epsilon$ was estimated in Ref.~\cite{Damour1996} with only binary-pulsar measurements available at the time; this constraint could be improved by including more recent observations.

\begin{table}[t]
\begin{ruledtabular}\caption{Constraints on the weak-field parameters in Eqs.~\eqref{eq:GPPN}--\eqref{eq:lambda2PPN} set by solar-system and binary-pulsar observations. As discussed in the text, we set $\phi_0=1$ for simplicity.}\label{table:constraints}
\begin{tabular}{C{.2\columnwidth}C{.24\columnwidth}C{.15\columnwidth}}
Parameter&Constraint&Reference\\
\hline
$\gamma_\text{PPN}-1$&$2.3\times 10^{-5}$&\cite{Bertotti2003}\\
$\beta_\text{PPN}-1$&$7.8\times 10^{-5}$&\cite{Antia2007,Will2014}\\
$\epsilon$&$7\times 10^{-2}$&\cite{Damour1996}\\
\hline
$G-1$&$1.2\times 10^{-5}$&\\
$\zeta$&$1.2\times 10^{-5}$&\\
$\bar\lambda_1$&$1.6\times 10^{-4}$&\\
$\bar\lambda_2$&$8.8\times 10^{-7}$&\\
\end{tabular}
\end{ruledtabular}
\end{table}

The current experimental constraints on the strong-field parameters $s_A,s'_A,s''_A$ are not as restrictive. The best limits on neutron-star sensitivities come from timing measurements of pulsar-white-dwarf binaries~\cite{Bhat2008, Antoniadis2010,Freire2012,Antoniadis2013}; white dwarfs are expected to have negligible sensitivity, so the magnitude of dipolar emission is dictated entirely by the sensitivity of the neutron star. Constraints are typically given in terms of the scalar charge $\alpha_A$, related to the sensitivity by
\begin{align}\label{eq:Chargedef}
\alpha_A=&\frac{1-2s_A}{\sqrt{3+2\omega_0}}.
\end{align}

Amongst known pulsar-white-dwarf binaries used to constrain scalar-tensor theories, PSR J0348+0432 hosts the most massive neutron star~\cite{Antoniadis2013}. The constraints on the scalar dipole reported in Ref.~\cite{Antoniadis2013} provide an estimate for the maximum scalar charge that this neutron star can have $|\alpha_A|\lesssim 6\times 10^{-3}$. Extending these data to an absolute bound on the charge of \emph{any} neutron star requires the assumption of a particular choice of $\omega(\phi)$ and equation of state. Working within one of the most popular classes of scalar-tensor theories \cite{Esposito-Farese1993} and selecting certain realistic equations of state, one can produce a global constraint of $|\alpha_A|\lesssim 10^{-2}$ \cite{Antoniadis2013,Wex2014}. However, it is conceivable that other theories and/or equations of state allow neutron stars to acquire large scalar charges of $\alpha_A\sim 1$ via the process of spontaneous scalarization \cite{Esposito-Farese1993} while satisfying all current experimental constraints.

Because the weak-field constraints leave $\omega_0\sim1/(2 \zeta)$ unbounded, no absolute bound can be placed on $s_A$. To our knowledge, no constraints have been placed on $s'_A$ and $s''_A$ either; for neutron stars, these higher derivatives can be orders of magnitude larger than $s_A$ (for example, see Fig. 3 of Ref.~\cite{Sennett2016}).

Excluding the possibility of spontaneous scalarization, the constraints on weak-field and strong-field parameters ensure that dipole radiation is suppressed in viable scalar-tensor theories, as can be shown by comparing the relative size of the $-1$PN and Newtonian order flux, given in Refs. \cite{Damour1992,Mirshekari2013} and repeated in Eq.~\eqref{eq:fluxDDsplit} below. Despite entering at higher PN order, the next-to-leading order term overpowers the leading-order term when
\begin{align}\label{eq:thresholdFreq}
1\lesssim \left(\frac{24}{5 \zeta \calS_-^2}\right)\left(G \alpha M \pi f \right)^{2/3},
\end{align}
where, for simplicity, we have dropped all terms that are not of order $\calO(\zeta^{-1})$ and introduced the scalar dipole
\begin{align}
\calS_-\equiv&-\alpha^{-1/2}\left(s_1-s_2\right).
\end{align}

Given the experimental constraints on $\zeta$ and $\calS_-$, this threshold is reached at frequencies ${f\lesssim100 \mu\text{Hz}}$ in binary neutron star or neutron-star stellar-mass-black-hole systems, and at frequencies ${f\lesssim5\mu\text{Hz}}$ in neutron-star intermediate-mass-black-hole systems. Following this argument, ground- and space-based GW detectors would only observe binary systems whose inspiral is driven by the next-to-leading order flux.\footnote{Unlike the class of scalar-tensor theories considered here, there are alternative theories in which binary black holes can emit dipolar radiation (e.g., dilatonic Einstein-Gauss-Bonnet, dynamical Chern-Simons, etc.). Given the relatively weak constraints on dipolar radiation in vacuum spacetimes (compared to those from binary pulsars observations), we note that space-based detectors or pulsar timing arrays could, in principle, observe binary black holes driven by dipolar flux. As discussed below, the GW signal from such systems has a distinct structure from that in GR.} On the other hand, the evolution of binary pulsars could be dominated by dipolar emission. Binary systems that undergo dynamical scalarization may also be exempt from this verdict, as these systems dynamically generate large scalar charges that can substantially enhance dipolar emission~\cite{Barausse2013,Shibata2014}.
%\footnote{In Ref.~\cite{Sennett2016}, the authors argue that PN approximation breaks down as dynamical scalarization occurs, but that a straightforward resummation of PN results can provide an accurate waveform model valid in this regime.}

Because non-perturbative scalarization phenomena have not been entirely ruled out, we compute below the gravitational waveform both for systems in which dipolar radiation is dominant and for those in which quadrupolar radiation is dominant. For conciseness, we refer to the former class of systems as \emph{dipole driven} (DD) and the latter class as \emph{quadrupole driven} (QD).

\subsection{Detector response}
We consider the response of a laser interferometer at spatial coordinates $\v{X}$ generated to an incident GW produced by a distant binary system of size $d$, where $R\equiv|\v{X}|\gg d$. We assume that far from the binary, the metric and scalar field approach the Minkowski metric $\eta^{\mu \nu}$ and a constant background value $\phi_0$, respectively, at a rate $\sim R^{-1}$. Let $\hat\phi\equiv \phi/\phi_0$ be the normalized scalar field. We introduce the conformally transformed metric
\begin{align}
\tilde{g}_{\mu \nu}&\equiv \hat\phi g_{\mu \nu},\label{eq:Defgtilde}
\end{align}
and the gravitational field\footnote{Note that in Ref.~\cite{Bliving} and the references therein, the metric perturbation is defined with an overall minus sign relative to the definition given here.}
\begin{align}
h^{\mu \nu}\equiv  \eta^{\mu \nu} - \sqrt{-\tilde{g}}\tilde{g}^{\mu \nu}. \label{eq:Defh}
\end{align}
The metric at the detector takes the form
\begin{align}
g_{\mu \nu}=&\eta_{\mu \nu}  + h_{\mu \nu} - \frac{1}{2}h\eta_{\mu \nu}-\Psi \eta_{\mu \nu}+\calO\left(R^{-2}\right),\label{eq:DefgFZ}
\end{align}
where $h\equiv \eta_{\mu \nu} h^{\mu \nu}$ is the trace of $h^{\mu \nu}$ and $h_{\mu \nu}\equiv \eta_{\mu \alpha} \eta_{\nu \beta} h^{\alpha \beta}$ is lowered using the Minkowski metric. Gravitational-wave detectors use laser interferometry to measure the separation between mirrors; we treat these mirrors as test masses. Assuming that the distance between mirrors is smaller than the wavelength of the incident GWs and that the mirrors move slowly, the separation between the mirrors obeys
\begin{align}
\ddot{\xi}^{i}=-R_{0i0j}\xi^{j},
\end{align}
where $i,j=1,2,3$ are spatial indices. Working at leading order in $h^{\mu \nu}$ and $\Psi$, the Riemann tensor is calculated from Eq.~\eqref{eq:DefgFZ}
\begin{align}
R_{0i0j}= -\frac{1}{2}\ddot{h}^{i j}_\text{TT}-\frac{1}{2}\ddot{\Psi}\left(\hat{N}^i \hat{N}^j-\delta^{ij}\right),\label{eq:RiemannTensor}
\end{align}
where $\uv{N}\equiv \v{X}/R$ and $h^{ij}_\text{TT}$ is the transverse-traceless component of the gravitational field defined as
\begin{align}\label{eq:TTproj}
h^{ij}_\text{TT}=\left(P^{ip}P^{jq}-\frac{1}{2}P^{ij}P^{pq}\right)h^{pq},
\end{align}
where $P^{pq}=\delta^{pq}-\hat{N}^p \hat{N}^q$ is the transverse projection operator.

From Eq.~\eqref{eq:RiemannTensor}, we see that the GW signal contains a transverse-traceless mode (as in GR) characterized by the field $h^{\mu \nu}$. In scalar-tensor gravity, there is an additional transverse breathing mode produced by $\Psi$. Extracting this new GW polarization requires a network of detectors; see Ref. \cite{Will2014} and references therein for a discussion of the prospects of detecting GW polarizations absent in GR. We focus exclusively on $h^{\mu \nu}$ for the remainder of this work.

%%%%%%%%%%%%%%%%%%%%%%%%%%%%%%%%%%%%

\section{Dynamics for quasi-circular orbits}
\label{sec:dynamics}

In this section, we specialize the results of Ref.~\cite{Mirshekari2013} for the 2.5PN dynamics of binary systems to the case of quasi-circular orbits. Before proceeding, we establish some notation employed throughout this work. We denote the total mass of the system by $M = m_{1} + m_{2}$ and the symmetric and antisymmetric mass ratio by $\eta = m_{1}m_{2}/M^{2}$ and $\psi = (m_{1}-m_{2})/M$, respectively. We signify the symmetric and antisymmetric combinations of parameters given in Table~\ref{table:parameters} by
\begin{subequations}
\begin{align}
\tau_+\equiv&\frac{1}{2}\left(\tau_1+\tau_2\right),\\
\tau_-\equiv&\frac{1}{2}\left(\tau_1-\tau_2\right),
\end{align}
\end{subequations}
and in addition to $\calS_-$ above, we also define
\begin{subequations}
\begin{align}
\calS_+\equiv&\alpha^{-1/2}\left(1-s_1-s_2\right).
%\calS_-\equiv&-\alpha^{-1/2}\left(s_1-s_2\right).
\end{align}
\end{subequations}

To describe the system's dynamics, we denote the orbital separation by $\v{x} = r \v{n}$, the relative velocity by $\v{v} = \dot{\v{x}}$, and the acceleration by $\v{a} = \dot{\v{v}}$. We construct an orthonormal moving frame $(\v{n}, \gv{\lambda})$ and define the orbital frequency $\omega$ such that $\v{v} = \dot{r}\v{n}+r \omega \gv{\lambda}$. To avoid confusion, we note that certain variables are used to denote multiple quantities; for example, $\omega$ represents both the frequency and scalar coupling, while $\phi,\Psi$ are used for the phase and scalar field. The usage of each can be inferred from context.

Our analysis of binary systems on quasi-circular orbits begins with the 2.5PN equation of motion, given in the center-of-mass frame by
\begin{align}
	\v{a} =& -\frac{G M \alpha}{r^{2}} \v{n} \nn\\
	&+ \frac{G M \alpha}{r^{2}} \left[ \v{n} \left( A_{\rm 1PN} + A_{\rm 2PN} \right) + \dot{r} \v{v} \left( B_{\rm 1PN} + B_{\rm 2PN} \right) \right] \nn\\
	&+ \frac{8\eta}{5} \frac{ (G M \alpha)^{2}}{r^{3}} \left[ \dot{r}\v{n} \left( A_{\rm 1.5PN} + A_{\rm 2.5PN} \right) \right. \nn\\
	& \left. \qquad \qquad \qquad  - \v{v} \left( B_{\rm 1.5PN} + B_{\rm 2.5PN} \right) \right] \,.
\end{align}
where the expressions for $A_i, B_i$ can be found in Eqs.~(1.4)--(1.5) and (6.12)--(6.13) of Ref.~\cite{Mirshekari2013}. It will prove useful also to write the equations of motion in the generic form
\be \label{eq:EOMgeometric}
	\v{a} = (\ddot{r} -r \omega^{2}) \v{n} + (r\dot{\omega} + 2\dot{r}\omega) \gv{\lambda} \,.
\ee
The restriction of the dynamics to quasi-circular orbits follows the same procedure as in GR. For such orbits, the only departure from circular motion is induced by radiation reaction, which enters at 1.5PN order in scalar-tensor theories rather than the usual 2.5PN order in GR.  Expressed symbolically, we have $\dot{r}, \,\dot{\omega} = \calO(3)$ [instead of $\calO(5)$], while $\ddot{r} = \calO(6)$ [instead of $\calO(10)$].
%In this section, we assume the relative size of the PN terms follow their standard hierarchy; the case in which dipolar radiation is subdominant to quadrupolar radiation is considered in Sec.~\ref{sec:GRlimit}.

The first term in Eq.~\eqref{eq:EOMgeometric} determines the conservative sector of the dynamics at 1PN and 2PN order. The scalar product $\v{a}\cdot\v{n} = -r\omega^{2} + \calO(6)$ produces a relation between the orbital separation and frequency that generalizes Kepler's law. We introduce the PN parameters (recall that we work in units where $c=1$)
\begin{subequations}
\begin{align}
	\gamma_{\rm PN} &\equiv \frac{G M\alpha }{r} , \\
	x &\equiv \left( G M \alpha \omega \right)^{2/3},
\end{align}
\end{subequations}
which differ from their usual definition in GR by an additional factor $\alpha$.

At leading order, one obtains $r^{3}\omega^{2}  = Gm\alpha +\calO(2)$, or $x = \gamma_{\rm PN} (1+\calO(2))$. From there, solving order by order yields
\begin{widetext}
\begin{subequations}\label{eq:xgammacirc}
\begin{align}
x &= \gamma_{\rm PN} \left[1 + \gamma_{\rm PN}\left(\frac{2 \beta_{-} \psi }{3}-\frac{2 \beta_{+}}{3}-\frac{\gamma }{3}+\frac{\eta }{3}-1\right) + \gamma_{\rm PN}^2\left(\frac{8 \beta_{-}^2 \eta }{\gamma }+\frac{16 \beta_{-}^2 \eta }{9}-\frac{4 \beta_{-}^2}{9}+\frac{8 \beta_{-} \beta_{+} \psi }{9}-\frac{2 \beta_{-} \gamma  \psi }{9}+\frac{11 \beta_{-} \psi  \eta }{9} \right. \right. \nn\\
 & \left. \left. -\frac{4 \beta_{-} \psi }{3}-\frac{8 \beta_{+}^2 \eta }{\gamma }-\frac{4 \beta_{+}^2}{9}+\frac{2 \beta_{+} \gamma }{9}+\frac{7 \beta_{+} \eta }{9}+\frac{4 \beta_{+}}{3}-\frac{\gamma ^2 \eta }{6}+\frac{11 \gamma ^2}{36}+\frac{17 \gamma  \eta }{9}+\gamma +\frac{\psi  \delta_{-}}{3}+\frac{2 \psi  \chi_{-}}{3}-\frac{2 \delta_{+} \eta }{3}+\frac{\delta_{+}}{3} \right. \right. \nn\\
 & \left. \left. +\frac{2 \eta ^2}{9}+\frac{4 \eta  \chi_{+}}{3}+\frac{49 \eta }{12}-\frac{2 \chi_{+}}{3}+1\right) + \calO(6) \right] , \\
\gamma_{\rm PN} &= x \left[1 + x\left(-\frac{2 \beta_{-} \psi }{3}+\frac{2 \beta_{+}}{3}+\frac{\gamma }{3}-\frac{\eta }{3}+1\right) + x^2\left(-\frac{8 \beta_{-}^2 \eta }{\gamma }-\frac{16 \beta_{-}^2 \eta }{3}+\frac{4 \beta_{-}^2}{3}-\frac{8 \beta_{-} \beta_{+} \psi }{3}-\frac{2 \beta_{-} \gamma  \psi }{3}-\frac{\beta_{-} \psi  \eta }{3} \right. \right. \nn\\
 & \left. \left. -\frac{4 \beta_{-} \psi }{3}+\frac{8 \beta_{+}^2 \eta }{\gamma }+\frac{4 \beta_{+}^2}{3}+\frac{2 \beta_{+} \gamma }{3}-\frac{5 \beta_{+} \eta }{3}+\frac{4 \beta_{+}}{3}+\frac{\gamma ^2 \eta }{6}-\frac{\gamma ^2}{12}-\frac{7 \gamma  \eta }{3}+\frac{\gamma }{3}-\frac{\psi  \delta_{-}}{3}-\frac{2 \psi  \chi_{-}}{3}+\frac{2 \delta_{+} \eta }{3}-\frac{\delta_{+}}{3} \right. \right. \nn\\
 & \left. \left. -\frac{4 \eta  \chi_{+}}{3}-\frac{65 \eta }{12}+\frac{2 \chi_{+}}{3}+1\right) + \calO(6) \right] .
\end{align}
\end{subequations}
\end{widetext}

Having derived the reduction to quasi-circular orbits for the conservative dynamics up to 2PN order, we now turn our attention to the dissipative sector. Although only the leading-order radiation-reaction terms are needed to compute the 2PN relative order dynamics, we provide results up to 2.5PN for the sake of completeness. Inserting Eq.~\eqref{eq:xgammacirc} into the relation $\v{a}\cdot \gv{\lambda} = r \dot{\omega} + 2 \dot{r} \omega$ gives the following expressions for $\dot{r}$ and $\dot{\omega}$
\begin{subequations}\label{eq:rdotomegadot}
\begin{align}
	\dot{r} &= -\frac{8}{3} \zeta  \eta  \calS_{-}^2 x^2 -\frac{8}{3} \eta  \delta_{\rm RR} x^3 + \calO(7)\,,\\
	\dot{\omega} &= \frac{4 \zeta  \eta  \calS_{-}^2 x^{9/2}}{ G^2 M^2 \alpha^{2}}  +\frac{4 \eta  \delta_{\rm RR} x^{11/2}}{G^2 M^2 \alpha^2}  + \calO(7) \,,
\end{align}
\end{subequations}
where we have introduced
\begin{align}
	\delta_{\rm RR} \equiv &\frac{24}{5 }+2 \gamma-\frac{4 \zeta\beta_{-} \psi \calS_{-}^2 }{\gamma }+\frac{8\zeta \beta_{-} \psi \calS_{-}^2}{3}-\frac{7\zeta \eta \calS_{-}^2 }{6}\nn\\
	&+\frac{4 \zeta\beta_{+}\calS_{-}^2}{\gamma }-\frac{8 \zeta\beta_{+}\calS_{-}^2}{3}+\frac{2 \zeta\gamma \calS_{-}^2}{3}-\frac{\zeta\calS_{-}^2}{2} \nn\\
	&+\frac{4\zeta \beta_{-} \calS_{+}\calS_{-}}{\gamma}-\frac{4\zeta \beta_{+} \psi  \calS_{+}\calS_{-}}{\gamma  } .
\end{align}

For dipole-driven systems, the second term in Eq.~\eqref{eq:rdotomegadot} is much smaller than the first. Integrating this equation at leading order gives the evolution of the orbital separation and frequency
\begin{subequations}\label{eq:scalingromegaST}
\begin{align}
	r_\text{DD}(t) =& \left[  8 \eta \zeta \calS_{-}^{2} (G M \alpha)^{2} (t_{c}-t)\right]^{1/3} \left( 1 + \calO(2) \right) , \\
	\omega_\text{DD}(t) =& \left[ 8 \eta \zeta \calS_{-}^{2} G M \alpha (t_{c}-t)\right]^{-1/2} \left( 1 + \calO(2) \right) ,
\end{align}
\end{subequations}
where $t_{c}$ is the time of coalescence. In the quadrupole-driven regime, the first term in Eq.~\eqref{eq:rdotomegadot} is overpowered by the second. We delay a precise formulation of this limit until Sec. \ref{sec:flux}, but note that the evolution of the inspiraling orbit will take the same form as in GR, given at leading order by
\begin{subequations}\label{eq:scalingromegaGR}
\begin{align}
	r_{\rm GR}(t) =& \left[  \frac{256 (G M)^3 \eta}{5} (t_{c}-t)\right]^{1/4} \left( 1 + \calO(2) \right) , \\
	\omega_{\rm GR}(t) =& \left[  \frac{256 (G M)^{5/3} \eta}{5} (t_{c}-t)\right]^{-3/8} \left( 1 + \calO(2) \right) .
\end{align}
\end{subequations}
The difference in structure between Eqs.~\eqref{eq:scalingromegaST} and~\eqref{eq:scalingromegaGR} stems from radiation reaction entering at a different PN order in the two regimes.

%%%%%%%%%%%%%%%%%%%%%%%%%%%%%%%%%%%%

\section{Radiative coordinates and hereditary contributions}
\label{sec:hered}

Equipped with the leading-order evolution of the inspiral, we begin our computation of the 2PN order waveform. The waveform was derived for generic orbits in Ref.~\cite{Lang2014a}; schematically, these results are given by

\begin{align}\label{eq:hijstructure}
	h^{ij}_{\rm TT} &= \frac{2G(1-\zeta)M\eta}{R} \left[ Q^{ij} + P^{1/2}Q^{ij} + PQ^{ij} + P^{3/2}Q^{ij}_{\calN} \right. \nn\\
	&  \qquad\left. + P^{3/2}Q^{ij}_{\calC-\calN} + P^{2}Q^{ij}_{\calN} + P^{2}Q^{ij}_{\calC-\calN}  \right]_{\rm TT}\,,
\end{align}
where $\rm TT$ stands for the transverse-traceless projection given in Eq.~\eqref{eq:TTproj} and $P$ denotes the PN order of each term. The expressions for $P^n Q^{ij}$ are presented in Eq.~(7.2) of Ref.~\cite{Lang2014a}. These terms can be categorized as either \emph{instantaneous} or \emph{hereditary}: instantaneous terms depend only on the current state of the system, whereas hereditary terms take the form of integrals extending over the binary's entire history. In Eq.~\eqref{eq:hijstructure}, $P^{3/2}Q^{ij}_{\calC-\calN}$ and $ P^{2}Q^{ij}_{\calC-\calN}$ are hereditary, while the remaining terms are all instantaneous.

This section details the computation of the hereditary terms for systems on quasi-circular orbits. First, we re-express the waveform in a radiative coordinate system, in which the metric perturbation falls off as $\sim R^{-1}$. We then compute separately the contributions from so-called \emph{tail} and \emph{memory} terms.

\subsection{Radiative coordinates}
\label{subsec:radiativecoord}

We begin by transforming the results of Ref.~\cite{Lang2014a} into radiative coordinates. This reference employed harmonic coordinates $X = (t,\v{X})$, defined by the gauge condition $\partial_{\nu}h^{\mu\nu} = 0$; however, these coordinates are known to give rise to unwanted logarithms of $R$ in the far-zone expansion. As shown in Ref.~\cite{Blanchet87}, it is possible to build another set of coordinates $\bar{X} = (\bar{t},\bar{\v{X}})$, called radiative coordinates, in which these logarithms are eliminated and the metric perturbation $h^{ij}$ admits an expansion in powers of $\bar{R}^{-1}$. Here, we will follow the presentation of Ref.~\cite{BD92}, in which the construction of this coordinate system is explicitly written at quadratic order in the multipolar post-Minkowskian formalism \cite{BD86}. Note that our definition $h^{\mu\nu}$ in Eq.~\eqref{eq:Defgtilde} introduces a sign difference with respect to Ref.~\cite{BD92}.

Both hereditary pieces, $P^{3/2}Q^{ij}_{\calC-\calN}$ and $ P^{2}Q^{ij}_{\calC-\calN}$, contain integrals with a logarithmic kernel, known as tail terms. The logarithmic terms can be expressed as the second time derivative of the leading-order, linearized metric, as shown by Eq.~(2.28) of Ref.~\cite{BD92}. Written in terms of the retarded time $u=t-R/c$, these terms are given by

%The important point here is that the integrand of these logarithmic terms is simply the second time derivative of the leading order, linearized metric, as shown by Eq.~(2.28) of~\cite{BD92}. Explicitly, we have for the logarithmic terms, with $u=t-R/c$ the retarded time:
\begin{widetext}
\begin{subequations}\label{eq:lnterms}
\begin{align}
	\left( P^{3/2}Q^{ij}_{\calC-\calN}(u) \right)_{\ln} &= 2G(1-\zeta)M \int_{0}^{+\infty} \ud s \, \frac{\ud^{2}}{\ud t^{2}} Q^{ij}(u-s) \ln\left( \frac{s}{2R+s} \right) , \\
	\left( P^{2}Q^{ij}_{\calC-\calN}(u) \right)_{\ln} &= 2G(1-\zeta)M \int_{0}^{+\infty} \ud s \, \frac{\ud^{2}}{\ud t^{2}}P^{1/2}Q^{ij}(u-s) \ln\left( \frac{s}{2R+s} \right) .
\end{align}
\end{subequations}
\end{widetext}

Because we are only interested in the $R^{-1}$ piece of the waveform, we expand the logarithms according to
\be\label{eq:expandlnR}
	\ln\left( \frac{s}{2R+s} \right) = \ln \left( \frac{s}{2R} \right) + \calO\left( \frac{1}{R} \right) .
\ee
We define the radiative coordinates as $\bar{X}^{\alpha} = X^{\alpha} + \xi^{\alpha}$, with
\be
	\xi^{\alpha} = 2G(1-\zeta)M \delta^{\alpha}_{\ph{\alpha}0}\ln\left( \frac{R}{r_{0}} \right) ,
\ee
where we have introduced an arbitrary constant length scale $r_{0}$. The metric perturbation in these new coordinates takes the form
\be
	\bar{h}^{\alpha\beta} = \left[ h^{\alpha\beta} - \partial^{\alpha}\xi^{\beta} - \partial^{\beta}\xi^{\alpha} + \eta^{\alpha\beta} \partial_{\rho}\xi^{\rho} + \xi^{\mu}\partial_{\mu} h^{\alpha\beta}\right]_{X=\bar{X}} \,,
\ee
where we have kept only the relevant terms in~Eqs.~(2.36) and (2.37) of Ref.~\cite{BD92}. The first three terms describe the usual effect of a first order gauge transformation on the harmonic perturbation; their contribution will be eliminated by the $\rm TT$ projection. The last term combines with the lower boundary term of the integrals in Eq.~\eqref{eq:lnterms} to replace $R$ by the new constant $r_{0}$:
\begin{widetext}
\begin{subequations}\label{eq:lntermsradiative}
\begin{align}
	\left( \overline{P^{3/2}Q^{ij}_{\calC-\calN}}(\bar{u}) \right)_{\ln} &= 2G(1-\zeta)M \int_{0}^{+\infty} \ud s \, \frac{\ud^{2}}{\ud \bar{t}^{2}}\overline{Q ^{ij}} (\bar{u}-s) \ln\left( \frac{s}{2r_{0}} \right) + \calO \left(\frac{1}{\bar{R}^{2}} \right), \\
	\left( \overline{P^{2}Q^{ij}_{\calC-\calN}} (\bar{u}) \right)_{\ln} &= 2G(1-\zeta)M \int_{0}^{+\infty} \ud s \, \frac{\ud^{2}}{\ud \bar{t}^{2}}\overline{P^{1/2}Q^{ij}}(\bar{u}-s) \ln\left( \frac{s}{2r_{0}} \right) + \calO \left(\frac{1}{\bar{R}^{2}} \right) .
\end{align}
\end{subequations}
\end{widetext}
Since the transformation to radiative coordinates affects only the logarithmic terms, from here on, we drop the notation $\bar{X}$, using instead the ordinary notation $X$ to signify these new coordinates.

%%%%%%
\subsection{Tail contributions}
\label{subsec:tail}

The tail terms in the waveform arise from back-scattering of the waves on the curvature of spacetime. In the multipolar post-Minkowskian wave generation formalism of Refs.~\cite{BD86, BD92}, they appear as interactions between each multipole moment and the mass monopole of the system. In the DIRE formalism~\cite{Will1996,Pati2000} used in Refs.~\cite{Lang2014a,Lang2015a}, tail terms arise from wave-zone contributions to the integrals over the past light-cone of the observer. Recall that these terms take the form of an integral with a logarithmic kernel over the past history of the source.

Since we are only interested in the $R^{-1}$ part of the waveform, we can expand the logarithms as in Eq.~\eqref{eq:expandlnR}. Using Eq.~\eqref{eq:xgammacirc} to replace $\omega$ with $r$, the tail terms then take the generic form
\be\label{eq:tailintegral}
	I = \int_{0}^{+\infty} \ud s \, \frac{e^{in\varphi(t-s)}}{r^{p}(t-s)} \ln \left( \frac{s}{2r_{0}} \right)
\ee
where $n,p$ are integers and $\varphi$ is the orbital phase of the binary.

To evaluate Eq.~\eqref{eq:tailintegral}, we make use of the fact that the radiation-reaction timescale is much longer than the orbital period. It was shown in Ref.~\cite{BS93} that ignoring radiation reaction, i.e., approximating the binary orbit as circular (with constant radius and frequency), introduces an error in these integrals of order~$\calO(\ln c/c^{5})$ in GR. The same argument holds in scalar-tensor gravity, with the only difference being that the error is of order $\calO(\ln c / c^{3})$ for dipole-driven systems due to the different scaling of radiation reaction. Under this assumption that the frequency does not evolve with $s$, we write $\varphi(t-s) \simeq \varphi(t) - s\omega(t)$. One can then compute the resulting integrals by making use of the formula~\cite{BS93}
\be
	\int_{0}^{+\infty} \ud y\, e^{i\lambda y} \ln y = -\frac{1}{\lambda} \left[ \frac{\pi}{2}\mathrm{sgn}(\lambda) + i \left( \gamma_{\rm E} + \ln |\lambda| \right) \right],
\ee
where $\gamma_{\rm E}$ is the Euler-Mascheroni constant.

%%%%%%
\subsection{Memory contributions}
\label{subsec:memory}

Memory terms arise in the waveform as integrals of the product of multipoles without a logarithmic kernel over the history of the source~\cite{Chr91,BD92}. They can be separated into so-called DC terms, which are non-oscillatory and accumulate over the entire lifetime of the system,  and AC, oscillatory terms that, by contrast, depend only on the recent history of the source.

The computation of oscillatory memory terms is identical in GR and in scalar-tensor theories. On quasi-circular orbits, these terms have the structure
\be
	J = \int_{0}^{+\infty} \ud s\, \frac{e^{in\varphi(t-s)}}{r^{p}(t-s)} ,
\ee
with integers $n,p$. Thanks to the oscillatory factor $e^{in\varphi}$ in the integrand, it can again be shown (see, e.g., Ref.~\cite{ABIQ04}) that only the recent past contributes in the integral, so that one can approximate $r(t-s)\simeq r(t)$ and $\varphi(t-s)\simeq \varphi(t)-s\omega(t)$ with a negligible relative error of the same PN order as radiation reaction. For dipole-driven inspirals, the result is
\be
	J_{\rm DD} = \frac{1}{in} \left( \frac{r(t)^{3}}{G M\alpha} \right)^{1/2} \frac{e^{in\varphi(t)}}{r^{p}(t)} + \calO(3) ,
\ee
whereas for quadrupole-driven inspirals, one obtains
\be
	J_{\rm QD} = \frac{1}{in} \left( \frac{r(t)^{3}}{G M\alpha} \right)^{1/2} \frac{e^{in\varphi(t)}}{r^{p}(t)} +\calO(5) .
\ee
Note that the only difference between these two cases is the order of the remainders.

The non-oscillatory (DC) memory terms take the form
\be\label{eq:DCmemory}
	K = \int_{0}^{+\infty} \ud s\, \frac{1}{r^{p}(t-s)} .
\ee
Their computations in GR and scalar-tensor theory differ. Non-oscillatory terms are enhanced by the accumulation of the integrand over the long radiation-reaction timescale, an effect which formally {\it decreases} their PN order. The result depends here on the rate of evolution of the quasi-circular inspiral under radiation reaction. For dipole-driven systems, these DC memory terms formally appear at the 1.5PN order in the expression of the multipole moments,  but the integration over the radiation-reaction timescale (formally of $-$1.5PN order) pushes this contribution back to Newtonian order. Using the leading-order evolution of the quasi-circular inspiral given by Eq.~\eqref{eq:scalingromegaST}, one obtains
\be\label{eq:DCmemoryDD}
	K_{\rm DD} =  \frac{3r^{3-p}(t)}{8(p-3)\zeta\eta\calS_{-}^{2}(G M \alpha)^{2}}  + \calO(-1),
\ee
for $p>3$.

The contribution from non-oscillatory memory terms in quadrupole-driven systems is more difficult to compute. Following Eq.~\eqref{eq:thresholdFreq}, any system with non-zero scalar dipole will have been dominated by dipolar radiation at some point during its lifetime. The transition between the dipole-driven and quadrupole-driven regimes needs to be accommodated in the integral in Eq.~\eqref{eq:DCmemory}. Such a calculation goes beyond the scope of this work.

%%%%%%%%%%%%%%%%%%%%%%%%%%%%%%%%%%%%

\section{Balance equation and phase evolution}
\label{sec:flux}

Having computed all of the hereditary terms for quasi-circular orbits, one can use Eqs.~\eqref{eq:xgammacirc} to express the waveform entirely in terms of the instantaneous orbital phase $\varphi$ and frequency $\omega$ of the binary. We now need the evolution of these quantities at 2PN order to finish our calculation of the waveform. This level of accuracy cannot be achieved using only the dynamics of the binary presented in Sec.~\ref{sec:dynamics}. In place of the higher-order radiation-reaction force, we use the total energy flux $\calF$ and the balance equation
\be
\frac{\ud E}{\ud t}=-\calF,
\ee
which can be reformulated using  $\dot{\varphi} = \omega$ as
\be\label{eq:balance}
	\frac{\ud \varphi}{\ud x} = -\frac{1}{G M\alpha} x^{3/2} \frac{\ud E/\ud x}{\calF(x)}.
\ee

We calculate the energy for systems restricted to quasi-circular orbits by applying the results of Sec.~\ref{sec:dynamics} to those of Ref.~\cite{Mirshekari2013}. The energy measured in an arbitrary frame is given by Eq.~(6.4) of Ref.~\cite{Mirshekari2013}. After shifting to the center-of-mass frame with Eqs.~(6.9) and~(6.10) of Ref.~\cite{Mirshekari2013}, we reduce this expression to the case of quasi-circular orbits using Eqs.~\eqref{eq:xgammacirc} and~\eqref{eq:rdotomegadot} and obtain
\begin{widetext}
\begin{align}\label{eq:Ecirc}
	E_{\rm circ} &= -\frac{1}{2} M \eta  x \left[1 + x\left(-\frac{2 \beta_{-} \psi }{3}+\frac{2 \beta_{+}}{3}-\frac{2 \gamma }{3}-\frac{\eta }{12}-\frac{3}{4}\right) + x^2\left(-\frac{16 \beta_{-}^2 \eta }{\gamma }-\frac{16 \beta_{-}^2 \eta }{3}+\frac{4 \beta_{-}^2}{3}-\frac{8 \beta_{-} \beta_{+} \psi }{3}-\frac{4 \beta_{-} \gamma  \psi }{3}\right. \right. \nn\\
 & \left. \left. +\frac{\beta_{-} \psi  \eta }{3}-\beta_{-} \psi +\frac{16 \beta_{+}^2 \eta }{\gamma }+\frac{4 \beta_{+}^2}{3}+\frac{4 \beta_{+} \gamma }{3}-\frac{19 \beta_{+} \eta }{3}+\beta_{+}+\frac{\gamma ^2 \eta }{3}-\frac{19 \gamma ^2}{12}+\frac{11 \gamma  \eta }{3}-\frac{14 \gamma }{3}+\frac{\psi  \delta_{-}}{3}-\frac{4 \psi  \chi_{-}}{3}\right. \right. \nn\\
 & \left. \left. +\frac{4 \delta_{+} \eta }{3}+\frac{\delta_{+}}{3}-\frac{\eta ^2}{24}-\frac{8 \eta  \chi_{+}}{3}+\frac{19 \eta }{8}+\frac{4 \chi_{+}}{3}-\frac{27}{8}\right) \right] \,.
\end{align}
\end{widetext}

The total emitted energy flux, including both tensor and scalar contributions, was given for generic orbits in Ref.~\cite{Lang2015a} with the structure
\be\label{eq:fluxstructure}
	\calF = \calF_{-1} + \calF_{0} + \calF_{0.5,\calC} + \calF_{0.5,\calC-\calN} + \calF_{1} + \calO(3) ,
\ee
where the number in the index indicates the PN order of each term. The $-$1PN term comes from dipolar, scalar radiation and is responsible for the appearance of radiation-reaction effects at 1.5PN order. Note that while the flux we consider is given at 1PN (using the order-counting scheme from GR), it corresponds to 2PN relative order.

The individual terms in Eq.~\eqref{eq:fluxstructure} are given in the center-of-mass frame in Eq.~(6.8) of Ref.~\cite{Lang2015a}. The term $\calF_{0.5,\calC-\calN}$ includes a logarithmic hereditary term coming from the product of the leading-order term and a tail term (at 1.5PN relative order) in the scalar waveform. We calculate this tail contribution using the method detailed in Sec.~\ref{subsec:tail}. Then, as before, we use the results of Sec.~\ref{sec:dynamics} to compute the total energy flux for quasi-circular orbits $\calF(x)$ that will be given in Eqs.~\eqref{eq:fluxDDsplit} and~\eqref{eq:fluxdipnondipsplit} below.

Equipped with expressions for the binding energy $E(x)$ and the total energy flux $\calF(x)$ both at the 2PN relative order, we proceed to evaluate the orbital phasing of the binary using Eq.~\eqref{eq:balance}. Different approaches have been proposed in the literature to integrate the balance equation, differing by the choice of integration variables (time or frequency) and by the choice of either numerical integration or analytical integration of a re-expansion of Eq.~\eqref{eq:balance} (see, e.g., Ref.~\cite{Buonanno+09} for a definition and comparison of these so-called Taylor approximants). Our purpose here is not to compare these different approaches, but to examine the new contributions in the phasing that arise in scalar-tensor theories. We adopt a method (corresponding to the TaylorT2 approximant) that provides a result in analytic form: the ratio $-(\ud E/\ud x)/\calF(x)$ is re-expanded in $x$, truncated at relative 2PN order, and then integrated term by term. For this purpose, it will be convenient to introduce the notation
\be\label{eq:defrhox}
	\rho(x) \equiv - \frac{1}{G M \alpha} \frac{1}{\calF(x)} \frac{\ud  E}{\ud x} .
\ee
Care must be taken before re-expanding this ratio in the PN parameter $x$: we distinguish between the dipole-driven case in which the dipolar term $\calF_{-1}$  [defined in Eq.~\eqref{eq:fluxstructure}] dominates the denominator and the quadrupole-driven case wherein $\calF_{0}$ dominates due to the smallness of scalar-tensor parameters.

\subsection{The dipole-driven regime}\label{subsec:phasingDD}

We first consider systems whose inspiral is driven by dipolar radiation. As discussed in Sec.~\ref{sec:constraints}, this regime is reached by binaries with large separations (binary pulsars) or large scalar dipoles (spontaneously scalarized systems). Dynamically scalarized systems begin in the quadrupole-driven regime but then abruptly become dipole-driven at some point during their evolution; in principle, one must account for both stages when modeling their inspiral, but we will not pursue such a treatment here.\footnote{In Ref.~\cite{Sennett2016}, the authors argue that PN approximation breaks down as dynamical scalarization occurs, but that a straightforward resummation of PN results can provide an accurate waveform model valid in this regime.}  Factoring out the leading order dipolar flux in Eq.~\eqref{eq:fluxstructure}, we obtain
\begin{align}\label{eq:fluxDDsplit}
  \calF^{\rm DD} (x) =& \frac{4 \mathcal{S}_-^2 \zeta \eta^2 x^4}{3 G \alpha} \left[ 1  + f_{2}^{\rm DD} x + f_{3}^{\rm DD} x^{3/2} \right. \nn\\
  & \qquad\qquad\quad\left. + f_{4}^{\rm DD} x^{2} + \mathcal{O}(5)\right] ,
\end{align}
where explicit expressions for the coefficients $f_{n}^{\rm DD}$ are given in Eq.~\eqref{eq:coeffsfluxDD}. The leading order of the flux carries a factor $\calS_{-}^{2}$ characteristic of dipolar radiation.

In this case, we simply re-expand the ratio $\rho(x)$ in $x$ at 2PN relative order and obtain
\begin{align}\label{eq:structrhoxDD}
  \rho^{\rm DD} (x) =& \frac{3}{8 \mathcal{S}_-^2 \zeta \eta x^4} \left[ 1  +  \rho^{\rm DD}_{2} x^{} +  \rho^{\rm DD}_{3} x^{3/2} \right. \nn\\
  & \qquad\qquad\quad \left. + \rho^{\rm DD}_{4} x^{2} + \mathcal{O}(5)\right] ,
\end{align}
where the coefficients $\rho_{n}^{\rm DD}$ are given explicitly in Eq.~\eqref{eq:coeffsrhoDD}. By integrating Eq.~\eqref{eq:balance} term-by-term, the phasing is then given by
%\begin{widetext}
\begin{align}
	\varphi(x) =& -\frac{1}{4 \mathcal{S}_-^2 \zeta \eta x^{3/2}} \left[ 1  + 3 x^{} \rho^{\rm DD}_{2} - \frac{3}{2} x^{3/2} \ln x \rho^{\rm DD}_{3} \right. \nn\\
  & \qquad\qquad\qquad\quad \left. - 3 x^{2} \rho^{\rm DD}_{4} + \mathcal{O}(5)\right] ,
\end{align}
where we have dropped an arbitrary additive constant that can be fixed by specifying the value of the phase at a given frequency.
%\end{widetext}

\subsection{The quadrupole-driven regime}\label{subsec:phasingQD}

For quadrupole-driven systems, the flux should be expanded about the Newtonian-order term $\calF_0$ in Eq.~\eqref{eq:fluxstructure} rather than the leading-order $-1$PN term. To accomplish this reordering of the PN approximation, we expand the flux in the PN parameter $x$ and an additional parameter that describes the smallness of non-GR effects. There exists some flexibility in the choice of this second small parameter; the weak-field parameters listed in Table \ref{table:parameters} describe the smallness of scalar-tensor corrections in complementary ways, and these quantities appear in the waveform in several combinations (e.g., the binary parameters).

We adopt a prescription that generalizes the approach of Ref.~\cite{Will1994} to more generic scalar-tensor theories and to higher PN order. In the present quadrupole-driven case, we split the flux into pieces independent and dependent on the scalar dipole
\begin{align}
	\calF^{\rm QD}=\calF_{\text{non-dip}}+\calF_{\text{dip}},
\end{align}
where we have defined
\begin{align}
\calF_{\text{non-dip}}\equiv& \lim_{\calS_-\rightarrow0}\calF,\\
\calF_{\text{dip}}\equiv& \calF-\calF_{\text{non-dip}}.
\end{align}
We refer to $\calF_{\text{dip}}$ and $\calF_{\text{non-dip}}$  as the ``dipolar part'' and ``non-dipolar part'' of the flux, respectively. Note that these labels do not correspond precisely to the multipolar structure of the source; for example, $\calF_{\text{dip}}$ contains contributions from time derivatives of the scalar monopole and quadrupole. Instead, $\calF_\text{dip}$ represents the part of the flux that vanishes when ${s_1=s_2}$

We compute the phasing at first order in the small quantity $\calF_{\text{dip}}/\calF_{\text{non-dip}}$, employing the approximation
\begin{align}\label{eq:STperttprime}
	-\frac{\ud E/\ud x}{\mathcal{F}(x)} \simeq -\frac{\ud E/\ud x}{\mathcal{F}_\text{non-dip}(x)}\left(1-\frac{\calF_{\text{dip}}(x)}{\calF_{\text{non-dip}}(x)}\right)
\end{align}
in Eq.~\eqref{eq:defrhox}. Evaluating the right-hand side of this equation requires knowledge of $\calF_\text{dip}$ and $\calF_\text{non-dip}$ each at 2PN relative order.

We obtain for the dipolar and non-dipolar parts
\begin{subequations}\label{eq:fluxdipnondipsplit}
\begin{align}
  \calF_\text{non-dip} (x) =& \frac{32  \eta^2 \xi x^5}{5 G \alpha} \left[ 1  + f^\text{nd}_{2} x^{} + \mathcal{O}(3)\right] , \label{eq:fluxnondipsplit}\\
  \calF_\text{dip} (x) =& \frac{4 \mathcal{S}_-^2 \zeta \eta^2 x^4}{3 G \alpha} \left[ 1  + f^\text{d}_{2} x^{} + f^\text{d}_{3} x^{3/2} \right. \nn \\
  & \qquad\qquad\qquad \left. + f^\text{d}_{4} x^{2} + \mathcal{O}(5)\right] , \label{eq:fluxdipsplit}
\end{align}
\end{subequations}
where the coefficients $f_{n}^\text{nd}$ and $f_{n}^\text{d}$ can be found in Eqs.~\eqref{eq:coeffsfluxdip} and~\eqref{eq:coeffsfluxnondip} of Appendix~\ref{app:coeffs}.  The leading order dipolar part of the flux~\eqref{eq:fluxdipsplit} is the same as in Eq.~\eqref{eq:fluxDDsplit}. The leading order non-dipolar part~\eqref{eq:fluxnondipsplit} is simply the quadrupolar flux in GR with an additional factor of $\xi/\alpha$, where we have defined $\xi \equiv 1+ \gamma/2 + \zeta \calS_{+}^{2}/6$.

Note that because it enters at Newtonian order (rather than $-1$PN), the non-dipolar part of the flux is only known to 1PN relative order. A complete calculation of the phasing at 2PN relative order requires the 1.5PN and 2PN corrections to the non-dipolar flux. In place of these unknown terms, we use
\begin{align}
	\calF_\text{non-dip}=&\calF^{(\text{GR})}_{\rm 2PN}+\calF^{(\text{ST})}_\text{non-dip},
\end{align}
with
\begin{align}\label{eq:deff3f4}
	\calF^{(\text{ST})}_\text{non-dip} =& \calF^{(\text{ST})\text{1PN}}_\text{non-dip} \nn\\ 
	&+ \frac{32 \eta^2 x^{5}}{5 G \alpha} \xi \left[ f^{\rm ST}_{3} x^{3/2}  + f^{\rm ST}_{4} x^{2} +\calO(5) \right].
\end{align}
In the above, $\calF^{(\text{GR})}_{\rm 2PN}$ is the PN expanded flux in GR up to 2PN order, with the natural replacement $G_*\rightarrow G\alpha$. The first term in Eq.~\eqref{eq:deff3f4} denotes the known contributions to the non-dipolar flux that only arise in scalar-tensor theories, which can be obtained by subtracting the GR terms from~\eqref{eq:fluxnondipsplit}. We introduce the unknown coefficients $f^{\rm ST}_{3}$ and $f^{\rm ST}_{4}$ to represent our ignorance of the new scalar-tensor contributions at 1.5PN and 2PN order. In the quadrupole-driven context, experimental constraints on the weak-field parameters imply that these contributions should be much smaller than the 2PN GR terms. Moreover, these terms are doubly suppressed in the second term of Eq.~\eqref{eq:STperttprime} because $\calF_\text{dip}$ is already of the first order in the small scalar-tensor coefficients. We will keep these unknown coefficients throughout our calculation for completeness.

%We argue that these unknown terms should have minimal impact on the overall GW signal based on the strong experimental constraints on the weak-field parameters. We see that the second term in Eq.~\eqref{eq:STperttprime} can be rewritten as
%\begin{align}
%\frac{\calF_{\text{dip}}}{\calF_{\text{non-dip}}}=\frac{\calF_{\text{dip}}}{\calF^{(\text{GR})}_\text{2PN}(1+\calF^{(\text{ST})}_{\text{non-dip}}/\calF^{(\text{GR})}_\text{2PN})},
%\end{align}
%demonstrating that these 1.5 and 2PN terms in $\calF^{(\text{ST})}_{\text{non-dip}}$ are doubly suppressed by the smallness of new scalar-tensor terms relative to those found in GR. However, for completeness we will keep the unknown coefficients $f^{\rm ST}_{3}$, $f^{\rm ST}_{4}$ throughout our calculation.

We repeat the computation of the phasing from Sec.~\ref{subsec:phasingDD} but using the approximation~\eqref{eq:STperttprime}. We write $\rho(x) = \rho_\text{non-dip}(x) + \rho_\text{dip}(x)$, where we have defined
\begin{subequations}\label{eq:rhodecompdef}
\begin{align}
	\rho_\text{non-dip}(x) \equiv& -\frac{1}{G M \alpha}\frac{1}{\mathcal{F}_\text{non-dip}(x)} \frac{\ud E}{\ud x}, \\
	\rho_\text{dip}(x) \equiv& \frac{1}{G M \alpha}\frac{\calF_{\text{dip}}(x)}{\mathcal{F}_\text{non-dip}(x)^{2}} \frac{\ud E}{\ud x},
\end{align}
\end{subequations}
which can be expanded in the form

\begin{subequations}
\begin{align}
	\rho_\text{non-dip}(x) =& \frac{5}{64 x^5 \eta \xi} \left[ 1  + \rho^{\rm nd}_{2} x + \rho^{\rm nd}_{3} x^{3/2} \right. \nn\\
  & \qquad\qquad \left. + \rho^{\rm nd}_{4} x^{2} + \mathcal{O}(5)\right] , \label{eq:structrhoxnondip} \\
	\rho_\text{dip}(x) =& - \frac{25 \mathcal{S}_-^2 \zeta}{1536 x^{6} \eta \xi^2} \left[ 1  + \rho^{\rm d}_{2} x + \rho^{\rm d}_{3} x^{3/2} \right. \nn\\
  & \qquad\qquad\qquad \left. + \rho^{\rm d}_{4} x^{2} + \mathcal{O}(5)\right] . \label{eq:structrhoxdip}
\end{align}
\end{subequations}
The expressions for the coefficients $\rho^{\rm nd}_{n} $, $\rho^{\rm d}_{n} $ are given in Eqs.~\eqref{eq:coeffsrhonondip} and~\eqref{eq:coeffsrhodip} in Appendix~\ref{app:coeffs}. Using the decomposition in Eq.~\eqref{eq:rhodecompdef}, we integrate Eq.~\eqref{eq:balance} and obtain the phase evolution
\begin{align}
	\varphi(x)=\varphi_\text{non-dip}(x)+\varphi_\text{dip}(x),
\end{align}
with
%\begin{widetext}
\begin{subequations}
\begin{align}
	\varphi_\text{non-dip}(x) =& -\frac{1}{32 x^{5/2}\eta \xi} \left[ 1  + \frac{5}{3}\rho^{\rm nd}_{2} x + \frac{5}{2} \rho^{\rm nd}_{3} x^{3/2}\right. \nn\\
  & \qquad\qquad\qquad \left. + 5 \rho^{\rm nd}_{4} x^{2} + \mathcal{O}(5) \vphantom{\frac{1}{2}}\right] , \label{eq:phixstructurenondip} \\
	\varphi_\text{dip}(x) =& \frac{25 \mathcal{S}_-^2 \zeta}{5376 x^{7/2} \eta \xi^2} \left[ 1  + \frac{7}{5}\rho^{\rm d}_{2} x + \frac{7}{4} \rho^{\rm d}_{3} x^{3/2}\right. \nn\\
  & \qquad\qquad\qquad\quad \left. + \frac{7}{3} \rho^{\rm d}_{4} x^{2} + \mathcal{O}(5)\right] , \label{eq:phixstructuredip}
\end{align}
\end{subequations}
where we have ignored an arbitrary additive constant phase.
%\end{widetext}

%%%%%%%%%%%%%%%%%%%%%%%%%%%%%%%%%%%%

\section{Spin-weighted spherical modes of the waveform}
\label{sec:modes}

Combining the results of the previous sections, we present the gravitational waveform in a convenient form for use with GW detectors. First, we decompose the waveform $h_{ij}^{\rm TT}$ as given in Eq.~\eqref{eq:hijstructure} into its plus and cross polarizations. We introduce the spherical coordinates $(R,\Theta,\Phi)$ in the center-of-mass frame and define the usual orthonormal triad $\{\uv{N},\uv{P},\uv{Q}\}$ where $\uv{N}=\v{e}_R$, and $\uv{P}$ and $\uv{Q}$ lie along the major and minor axes, respectively, of the projection of the orbital plane onto the plane of the sky. The plus and cross polarizations of the waveform are defined as the projections
%\footnote{Note that different choices for the vectors $\uv{P},\uv{Q}$ lead to sign changes for the polarizations $h_{+},h_{\times}$.}
\begin{subequations}
\begin{align}
	h_{+} =& \frac{1}{2} \left( \hat P_{i} \hat P_{j} - \hat Q_{i}\hat Q_{j} \right) h_{ij}^{\rm TT} \,, \\
	h_{\times} =& \frac{1}{2} \left( \hat P_{i} \hat Q_{j} + \hat Q_{i}\hat P_{j} \right) h_{ij}^{\rm TT} \,.
\end{align}
\end{subequations}
We then decompose the waveform into spin-weighted spherical harmonics according to~\cite{Thorne80}
\be
	h_{+} - i h_{\times} = \sum\limits_{\ell\geq 2} \sum\limits_{m=-\ell}^{\ell} {}_{-2}Y_{\ell m} (\Theta, \Phi) h_{\ell m} ,
\ee
where the coefficients $h_{\ell m}$ are the spin-weighted spherical modes that we wish to compute.

We introduce, as in GR, a convenient new orbital phase variable that allows us to formally absorb the logarithms appearing in the polarizations $h_{+}, h_{\times}$:
\be\label{eq:defpsi}
	\phi \equiv \varphi - \frac{2(1-\zeta)}{\alpha} x^{3/2} \left[ \ln \left( 4 r_{0} \omega \right) + \gamma_{\rm E} - \frac{11}{12}\right],
\ee
where $r_{0}$ is the length scale associated with the transformation to radiative coordinates introduced in Sec.~\ref{subsec:radiativecoord}. This definition differs from its GR counterpart\footnote{Note that in the notation of Ref.~\cite{Bliving} and references therein, this redefined phase is denoted by $\psi$. We instead use $\phi$ to avoid confusion with our $\psi = (m_{1} - m_{2})/M$.}~\cite{BDIWW95} by only a factor of $(1-\zeta)/\alpha$. Note that the difference between $\phi$ and $\varphi$ is of at least 3PN relative order because the leading-order term in the phase is formally $\calO(-3)$ in dipole-driven systems and $\calO(-5)$ in quadrupole-driven systems. Given that we control the phasing of the binary only at 2PN relative order, we can ignore this correction.

For the mode amplitudes, we adopt the notation
\be\label{eq:defHhatlm}
	h_{\ell m} \equiv \frac{2G M (1-\zeta) \eta x}{R } \sqrt{\frac{16\pi}{5}}\hat{H}_{\ell m} e^{-i m \phi} ,
\ee
where the appropriate phase factor is scaled out for each mode as well as the leading order amplitude of the 22 mode, which differs from its value in GR by only a factor of $1-\zeta$.

Because we consider only non-spinning binaries, and consequently, those on planar orbits, the modes obey the symmetry relation
\be
	h_{\ell m} = (-1)^{\ell} h_{\ell, -m}^{*} \,.
\ee
Thus, one needs only the modes with $m \geq 0$ to specify the waveform. Combining the results of the previous sections, we obtain at 2PN order for the quantities $\hat{H}_{\ell m}$\footnote{Recall that our definition for $h^{\mu\nu}$~\eqref{eq:Defgtilde} introduces a sign difference relative to the results summarized in Ref.~\cite{Bliving}.}:

\allowdisplaybreaks
\begin{subequations}\label{eq:Hhatlm}
\begin{widetext}
\begin{align}
	\hat{H}_{2,2} &= 1 + x\left(\frac{4 \beta_{-} \psi }{3}-\frac{4 \beta_{+}}{3}-\frac{2 \gamma }{3}+\frac{55 \eta }{42}-\frac{107}{42}\right) + x^{3/2}\left(-\frac{2 \pi  \zeta }{\alpha }+\frac{2 \pi }{\alpha }-\frac{3}{2} i \zeta  \eta  \calS_{-}^2-\frac{1}{3} i \zeta  \calS_{-}^2+\frac{1}{3} i \zeta  \calS_{+}^2\right) \nn\\
 & + x^2\left(\frac{16 \beta_{-}^2 \eta }{\gamma }+\frac{16 \beta_{-}^2 \eta }{3}-\frac{4 \beta_{-}^2}{3}+\frac{8 \beta_{-} \beta_{+} \psi }{3}+\frac{19 \beta_{-} \psi  \eta }{7}-\frac{113 \beta_{-} \psi }{63}-\frac{16 \beta_{+}^2 \eta }{\gamma }-\frac{4 \beta_{+}^2}{3} +\frac{23 \beta_{+} \eta }{7} \right. \nn\\
 & \left. +\frac{113 \beta_{+}}{63}-\frac{\gamma ^2 \eta }{3}+\frac{5 \gamma ^2}{12}-\frac{74 \gamma  \eta }{21}-\frac{\gamma }{21}+\frac{\psi  \delta_{-}}{3}+\frac{4 \psi  \chi_{-}}{3}-\frac{4 \delta_{+} \eta }{3}+\frac{\delta_{+}}{3}+\frac{2047 \eta ^2}{1512}+\frac{8 \eta  \chi_{+}}{3}-\frac{1069 \eta }{216}-\frac{4 \chi_{+}}{3}-\frac{2173}{1512}\right) \\
	\hat{H}_{2,1} &= \frac{1}{3} i \psi  \sqrt{x} \left[1 + x\left(2 \beta_{-} \psi -2 \beta_{+}+\frac{\gamma }{2}+\frac{5 \eta }{7}-\frac{17}{28}\right) \right. \nn\\
 & \left. + x^{3/2}\left(-\frac{\pi  \zeta }{\alpha }+\frac{i \zeta }{2 \alpha }+\frac{i \zeta  \ln (16)}{2 \alpha }+\frac{\pi }{\alpha }-\frac{i}{2 \alpha }-\frac{i \ln (16)}{2 \alpha }-\frac{4}{3} i \zeta  \eta  \calS_{-}^2-\frac{4 i \zeta  \eta  \calS_{-} \calS_{+}}{3 \psi }\right) \right] \\
	\hat{H}_{3,3} &= -\frac{3}{4} i \sqrt{\frac{15}{14}} \psi  \sqrt{x} \left[\vphantom{-\frac{3 \pi  \zeta }{\alpha }}1 + x\left(2 \beta_{-} \psi -2 \beta_{+}-\gamma +2 \eta -4\right) \right. \nn\\
 & \left. + x^{3/2}\left(-\frac{3 \pi  \zeta }{\alpha }+\frac{21 i \zeta }{5 \alpha }-\frac{6 i \zeta  \ln \left(\frac{3}{2}\right)}{\alpha }+\frac{3 \pi }{\alpha }-\frac{21 i}{5 \alpha }+\frac{6 i \ln \left(\frac{3}{2}\right)}{ \alpha }-\frac{8}{9} i \zeta  \eta  \calS_{-}^2-\frac{3}{10} i \zeta  \calS_{-}^2+\frac{8 i \zeta  \eta  \calS_{-} \calS_{+}}{9 \psi }+\frac{3}{10} i \zeta  \calS_{+}^2\right) \right] \\
	\hat{H}_{3,2} &= \frac{x}{54 \sqrt{35}} \left[ 90-270 \eta + x\left(-720 \beta_{-} \psi  \eta +240 \beta_{-} \psi +720 \beta_{+} \eta -240 \beta_{+}-365 \eta ^2+725 \eta -193\right) \right] \\
	\hat{H}_{3,1} &= \frac{i \psi  \sqrt{x}}{12 \sqrt{14}} \left[1 + x\left(2 \beta_{-} \psi -2 \beta_{+}-\gamma -\frac{2 \eta }{3}-\frac{8}{3}\right) \right. \nn\\
 & \left. + x^{3/2}\left(-\frac{\pi  \zeta }{\alpha }+\frac{7 i \zeta }{5 \alpha }+\frac{2i \zeta  \ln (2)}{\alpha }+\frac{\pi }{\alpha }-\frac{7 i}{5 \alpha }-\frac{2i \ln (2)}{ \alpha }-\frac{40}{3} i \zeta  \eta  \calS_{-}^2-\frac{1}{10} i \zeta  \calS_{-}^2+\frac{8 i \zeta  \eta  \calS_{-} \calS_{+}}{3 \psi }+\frac{1}{10} i \zeta  \calS_{+}^2\right) \right] \\
	\hat{H}_{4,4} &= \frac{4 x}{297 \sqrt{35}} \left[ 990 \eta -330 + x\left(2640 \beta_{-} \psi  \eta -880 \beta_{-} \psi -2640 \beta_{+} \eta +880 \beta_{+}-1320 \gamma  \eta +440 \gamma +2625 \eta ^2-6365 \eta +1779\right) \right] \\
	\hat{H}_{4,3} &= \frac{9 i \psi  (2 \eta -1) x^{3/2}}{4 \sqrt{70}} \\
	\hat{H}_{4,2} &= -\frac{1}{63} \sqrt{5} x \left[3 \eta -1 + x\left(8 \beta_{-} \psi  \eta -\frac{8 \beta_{-} \psi }{3}-8 \beta_{+} \eta +\frac{8 \beta_{+}}{3}-4 \gamma  \eta +\frac{4 \gamma }{3}+\frac{19 \eta ^2}{22}-\frac{805 \eta }{66}+\frac{437}{110}\right) \right]\\
\end{align}
\end{widetext}
\begin{align}
\hat{H}_{4,1} &= -\frac{i \psi  (2 \eta -1) x^{3/2}}{84 \sqrt{10}} \\
\hat{H}_{5,5} &= -\frac{625 i \psi  (2 \eta -1) x^{3/2}}{96 \sqrt{66}} \\
\hat{H}_{5,4} &= -\frac{32 \left(5 \eta ^2-5 \eta +1\right) x^2}{9 \sqrt{165}} \\
\hat{H}_{5,3} &= \frac{9}{32} i \sqrt{\frac{3}{110}} \psi  (2 \eta -1) x^{3/2} \\
\hat{H}_{5,2} &= \frac{2 \left(5 \eta ^2-5 \eta +1\right) x^2}{27 \sqrt{55}} \\
\hat{H}_{5,1} &= -\frac{i \psi  (2 \eta -1) x^{3/2}}{288 \sqrt{385}} \\
\hat{H}_{6,6} &= \frac{54 \left(5 \eta ^2-5 \eta +1\right) x^2}{5 \sqrt{143}} \\
\hat{H}_{6,5} &= 0 \\
\hat{H}_{6,4} &= -\frac{128}{495} \sqrt{\frac{2}{39}} \left(5 \eta ^2-5 \eta +1\right) x^2 \\
\hat{H}_{6,3} &= 0 \\
\hat{H}_{6,2} &= \frac{2 \left(5 \eta ^2-5 \eta +1\right) x^2}{297 \sqrt{65}} \\
\hat{H}_{6,1} &= 0 ,
\end{align}
\end{subequations}
where we omit the common remainder $\calO(5)$ for all modes.

The $h_{\ell m}$ modes with $m=0$ correspond to non-oscillatory memory terms. As discussed in Sec.~\ref{subsec:memory}, even systems presently driven by quadrupolar radiation will have undergone a dipole-driven phase in the distant past, which complicates the calculation of these DC memory terms. Hence, we limit ourselves to the dipole-driven case and use Eq.~\eqref{eq:DCmemoryDD}. Working at Newtonian order, the only non-zero mode is $h_{20}$, which reads
\be
		\hat{H}_{2,0}^{\rm DD} = \frac{1}{4 \sqrt{6}} + \calO(2) .
\ee
%Recall that the computation of the DC memory term in Eq.~\eqref{eq:H20term} was only done for dipolar-radiation driven systems; the corresponding term for quadrupolar-radiation driven systems remains unknown. All other results are valid for systems in either regime.

%%%%%%%%%%%%%%%%%%%%%%%%%%%%%%%%%%%%

\section{Stationary phase approximation}
\label{sec:spa}
In this section, we compute the Fourier transform of the gravitational waveform using the stationary phase approximation (SPA). This technique is only applicable to oscillatory modes; we do not consider the $m=0$ modes here.

We adopt the following convention for the Fourier transform of a function $g$:
\be\label{eq:FourierTransform}
	\tilde{g} (f) \equiv \int_{-\infty}^{+\infty} \ud t \, e^{+2i\pi f t} g(t) .
\ee
Note that this convention differs from the standard one, in which the argument of the exponential has a minus sign. Our convention ensures that modes proportional to $e^{-i m \varphi}$ with positive mode number $m$ and increasing orbital phase $\varphi$ have power in positive frequencies in the Fourier domain. Our results can be converted to the more common convention by taking $f\rightarrow-f$.

Combining the terms in Eq.~\eqref{eq:defHhatlm}, the $h_{\ell m}$ modes can be written as
\be\label{eq:hlmDecomposition}
	h_{\ell m}(t) = A_{\ell m}(t) e^{-im\varphi(t)} ,
\ee
where $A_{\ell m}$ is the (complex) amplitude. Note that we use $\varphi$ to describe the phase rather than $\phi$ defined in Eq.~\eqref{eq:defpsi} --- we ignore the 3PN correction $\phi-\varphi$, which can be thought of as a small phase correction to the amplitude at higher order than we work.

%We will also use the orbital frequency $\omega = \dot{\varphi}$. Note that we are free to allow the amplitudes $A_{\ell m}$ to be complex (through the factors $\hat{H}_{\ell m}$ given in []) and carry a phase on their own, provided they remain slowly variable. In this section, noting that the correction $\phi - \varphi$ given in [] is formally a 3PN correction in the phase, we will simply ignore it, although it could be straigtforwardly included as a small phase correction in the complex amplitudes $A_{\ell m}$. The conditions for the stationary phase approximation (SPA) to apply, $|\dot{A}_{\ell m}/A_{\ell m} |\ll \omega$, $|\dot{\omega} |\ll \omega^{2}$ and $|\dot{A}_{\ell m}/A_{\ell m} |^{2} \ll |\dot{\omega}| $, are well verified during the inspiralling phase of the binary~\cite{}.

The $h_{\ell m}$ modes for $m\neq 0$ are rapidly oscillatory, slowly chirping signals. Put more precisely, during the inspiral, the modes satisfy $|\dot{A}_{\ell m}/A_{\ell m} |\ll \omega$ and $|\dot{\omega} |\ll \omega^{2}$, which indicates that the SPA is applicable to the waveform~\cite{Thorne300}. Applying the Fourier transform~\eqref{eq:FourierTransform} to Eq.~\eqref{eq:hlmDecomposition} gives
\begin{subequations}
\begin{align}\label{eq:SPA}
	h^{\mathrm{SPA}}_{\ell m}(f) &= \calA_{\ell m} (f) e^{-i\Psi_{\ell m}(f)-i\pi/4}\,, \\
	\Psi_{\ell m}(f) &= m\varphi(t^{(m)}_{f}) - 2\pi f t^{(m)}_{f}\,, \label{eq:SPAphase} \\
	\calA_{\ell m} (f) &= A_{\ell m}(t^{(m)}_{f}) \sqrt{\frac{2\pi}{m\dot{\omega}(t^{(m)}_{f})}} ,\label{eq:SPAamp}
\end{align}
\end{subequations}
where $\omega=\dot\varphi$ and $t^{(m)}_{f}$ is defined implicitly as the time at which $m\omega(t^{(m)}_{f}) = 2\pi f$. Note the $m$-dependence of this time-to-frequency correspondence; at a given time, the different harmonics in the signal correspond to gravitational wave emission at different frequencies.

In keeping with the notation common in the literature, we introduce the new PN tracking parameter $v=x^{1/2}=(G M\alpha \omega)^{1/3}$, tied to the orbital frequency $\omega$. It is customary to introduce a similar notation for the frequency $f$ as $v_{f} = (\pi G M\alpha f)^{1/3}$. Since $v(t_{f}^{(m)}) = (2/m)^{1/3}v_{f}$, Eq.~\eqref{eq:SPAphase} can be rewritten as
\be \label{eq:Psivarphit}
	\Psi_{\ell m} (f) = m \left. \left(\varphi(v) - \frac{1}{G M \alpha} v^{3}t(v)\right) \right|_{v=(2/m)^{1/3}v_{f}} .
\ee

We compute the functions $\varphi(v)$ and $t(v)$ using a similar method to the phasing as in Eq.~\eqref{eq:balance}. From the balance equation~\eqref{eq:balance} we deduce
\begin{subequations}\label{eq:phitofv}
\begin{align}
	 \varphi(v) &= \varphi(v_{0}) - \frac{1}{G M\alpha} \int_{v_{0}}^{v}\ud v \, v^{3} \frac{\ud E/\ud v}{\mathcal{F}(v)} , \\
	 t(v) &= t(v_{0}) - \int_{v_{0}}^{v}\ud v \, \frac{\ud E/\ud v}{\mathcal{F}(v)} ,
\end{align}
\end{subequations}
where $v_{0}$ is related to the orbital frequency at some reference point in the evolution. Likewise, the factor entering the Fourier-domain amplitude~\eqref{eq:SPAamp} is computed using
\be\label{eq:omegadot}
	\frac{1}{\dot{\omega}} = -\frac{G M \alpha}{3 v^{2}} \frac{1}{\mathcal{F}(v)} \frac{\ud E}{\ud v} .
\ee

We evaluate Eqs.~\eqref{eq:phitofv} and~\eqref{eq:omegadot} using a prescription akin to that in Sec.~\ref{sec:flux} (corresponding now to the TaylorF2 approximant~\cite{Buonanno+09}): the expressions are re-expanded in $v$, truncated at relative 2PN order, and then integrated term by term. For the sake of compactness, we write $\varphi(v)$, $t(v)$, and $1/\dot{\omega}$ in terms of the expansion of the dimensionless ratio $\rho(x)$ introduced in Eq.~\eqref{eq:defrhox}, using
\be\label{eq:defrhov}
	- \frac{1}{  G M \alpha} \frac{1}{\mathcal{F}(v)} \frac{\ud E}{\ud v} = 2 v \rho(v^{2}) .
\ee

\subsection{The dipole-driven regime}\label{subsec:SPADD}

%We call $t_{0}$ and $\varphi_{0}$ the constants that appear when aggregating $\varphi(v_{0})$, $t(v_{0})$ with the terms coming from the lower boundary of the integrals. In terms of the coefficients $\rho_{n}$, we obtain

For the dipole-driven regime, we insert Eq.~\eqref{eq:defrhov} into Eq.~\eqref{eq:phitofv} using the expansion~\eqref{eq:structrhoxDD} and integrate, yielding
\begin{subequations}
\begin{align}
	\varphi^{\rm DD}(v) =& \frac{-1}{4 \calS_{-}^{2}\zeta \eta v^{3}} \left[ 1 + 3\rho_{2}^{\rm DD} v^{2} - 3 \rho_{3}^{\rm DD} v^{3} \ln v - 3 \rho_{4}^{\rm DD} v^{4} \right] , \\
	\frac{t^{\rm DD}(v)}{G M \alpha} =& \frac{-1}{8 \calS_{-}^{2} \zeta \eta v^{6}} \left[ 1 + \frac{3}{2}\rho_{2}^{\rm DD} v^{2} + 2\rho_{3}^{\rm DD} v^{3} + 3 \rho_{4}^{\rm DD} v^{4} \right] ,
\end{align}
\end{subequations}
where we have dropped the integration constants for now.
\begin{widetext}
Combining these two expressions gives the SPA phase~\eqref{eq:Psivarphit}
\begin{align} \label{eq:SPAphaserhocoeffs}
	\Psi_{\ell m}^{\rm DD} (f) =& - \frac{m}{8 \calS_{-}^{2} \zeta \eta v^{3}} \left[ 1 + \frac{9}{2}\rho_{2}^{\rm DD} v^{2} - 2\rho_{3}^{\rm DD} v^{3}(1+3\ln v) - 9 \rho_{4}^{\rm DD} v^{4} \right] +m \varphi_{0} - 2\pi f t_{0}\,,
\end{align}
where $v$ is evaluated at $v=(2/m)^{1/3}v_{f}$, and where we restored constants $t_{0}$ and $\varphi_{0}$ which are the sums of $\varphi(v_{0})$, $t(v_{0})$ with the terms from the lower boundary of the integrals. The coefficients $\rho_{n}^{\rm DD}$ are given in Eq.~\eqref{eq:coeffsrhoDD}.

Similarly, the complex amplitude is given by
\begin{align}
	\calA_{\ell m}^{\rm DD} (f) =& \frac{2G^2 (1-\zeta)M^{2}\alpha \pi \eta^{1/2}}{R \zeta^{1/2}|\calS_-|} \sqrt{\frac{4}{5}} \sqrt{\frac{2}{m}} \frac{\hat{H}_{\ell m}(v)}{v^{5/2}} \left[ 1  + \frac{1}{2}v^{2} \rho_{2}^{\rm DD} + \frac{1}{2}v^{3} \rho_{3}^{\rm DD} + \frac{1}{2}v^{4}\left(\rho_4^{\rm DD}-\frac{1}{4}(\rho_2^{\rm DD})^2\right)\right],
\end{align}
\end{widetext}
where $\hat{H}_{\ell m}(v)$ is given by Eq.~\eqref{eq:Hhatlm} with the replacement $x=v^{2}$ and, as before, $v$ is evaluated at $v=(2/m)^{1/3}v_{f}$.

Note that because $\hat{H}_{\ell m}$ are complex they can affect the phase of the waveform. In particular, they can carry an overall minus sign or factor $\pm i$, which should be included in the phase of the mode. In addition, higher-order terms can have a factor $\pm i$ differing from the one entering at leading order, which induce corrective phases. However, those phases turn out to be negligible, entering at higher PN order than the 2PN relative order we consider.

\subsection{The quadrupole-driven regime}
We follow the same treatment for the quadrupole-driven systems as laid out in Sec. \ref{subsec:phasingQD}: we split the flux into dipolar and non-dipolar parts, and expand $\rho(v)$ to first order in the ratio $\calF_\text{dip}/\calF_\text{non-dip}$ according to Eq.~\eqref{eq:STperttprime}. The first and second terms in Eq.~\eqref{eq:STperttprime} produce a non-dipolar and dipolar contribution, respectively, to the phase and amplitude of the SPA waveform.

The non-dipolar contribution to the phasing is constructed by inserting Eq.~\eqref{eq:defrhov} into the integrals in Eq.~\eqref{eq:phitofv} and using the expansion~\eqref{eq:structrhoxnondip}. Ignoring the integration constants for now, we find
\begin{widetext}
\begin{subequations}
\begin{align}
	\varphi^\text{non-dip}(v) =& -\frac{1}{32 v^{5}\eta \xi} \left[ 1  + \frac{5}{3}\rho^{\rm nd}_{2} v^{2} + \frac{5}{2} \rho^{\rm nd}_{3} v^{3} + 5 \rho^{\rm nd}_{4} v^{4} + \mathcal{O}(5)\right] , \label{eq:phivstructurenondip} \\
	\frac{t^\text{non-dip}(v)}{G M\alpha} =&  -\frac{5}{256 v^{8}\eta \xi} \left[ 1  + \frac{4}{3}\rho^{\rm nd}_{2} v^{2} + \frac{8}{5} \rho^{\rm nd}_{3} v^{3} + 2 \rho^{\rm nd}_{4} v^{4} + \mathcal{O}(5)\right] ,
\end{align}
\end{subequations}
and thus, the corresponding contribution to the Fourier-domain phase for the $h_{\ell m}$ mode is given by
\be
	\Psi_{\ell m}^\text{non-dip} (f) = m \left. \left[ -\frac{3}{256 v^{5}\eta \xi} \left( 1  + \frac{20}{9}\rho^{\rm nd}_{2} v^{2} + 4 \rho^{\rm nd}_{3} v^{3} + 10 \rho^{\rm nd}_{4} v^{4} \right)  \right] \right|_{v=(2/m)^{1/3}v_{f}}.  \label{eq:nonDipSPAphase}
\ee
Similarly, we use Eq.~\eqref{eq:structrhoxdip} to compute the contribution to the phasing from the dipolar energy flux
\begin{subequations}  \label{eq:DipSPAphase}
\begin{align}
	\varphi^\text{dip}(v) =& \frac{25 \mathcal{S}_-^2 \zeta}{5376 v^{7} \eta \xi^2} \left[ 1  + \frac{7}{5}\rho^{\rm d}_{2} v^{2} + \frac{7}{4} \rho^{\rm d}_{3} v^{3} + \frac{7}{3} \rho^{\rm d}_{4} v^{4} + \mathcal{O}(5)\right] , \label{eq:phivstructuredip} \\
	\frac{t^\text{dip}(v)}{G M\alpha} =& \frac{5 \mathcal{S}_-^2 \zeta}{1536 v^{10} \eta \xi^2} \left[ 1  + \frac{5}{4}\rho^{\rm d}_{2} v^{2} + \frac{10}{7} \rho^{\rm d}_{3} v^{3} + \frac{5}{3} \rho^{\rm d}_{4} v^{4} + \mathcal{O}(5)\right] ,\\
	\Psi_{\ell m}^\text{dip} (f) =& m \left. \left[ \frac{5 \mathcal{S}_-^2 \zeta}{3584 v^{7} \eta \xi^2} \left( 1  + \frac{7}{4}\rho^{\rm d}_{2} v^{2} + \frac{5}{2} \rho^{\rm d}_{3} v^{3} + \frac{35}{9} \rho^{\rm d}_{4} v^{4} \right)  \right] \right|_{v=(2/m)^{1/3}v_{f}}.
\end{align}
\end{subequations}
\end{widetext}

Combining these two pieces and restoring the constants $\varphi_{0}$ and $t_{0}$, the Fourier-domain phase is then simply
\be\label{eq:Psilmtotal}
	\Psi_{\ell m}^{\rm QD}(f) = \Psi_{\ell m}^\text{non-dip} (f) + \Psi_{\ell m}^\text{dip} (f) + m\varphi_{0} - 2\pi f t_{0} .
\ee
The coefficients $\rho^{\rm nd}_{n}$ and $\rho^{\rm d}_{n}$ are given in Eqs.~\eqref{eq:coeffsrhonondip} and~\eqref{eq:coeffsrhodip}. When restricted to Brans-Dicke theory, Eq.~\eqref{eq:Psilmtotal} reproduces the leading order deviation in the phase from GR derived in Ref.~\cite{Will1994} at order $\calO(1/\omega_\text{BD})$, which is equivalent to first order in $\calF_{\text{dip}}/\calF_{\text{non-dip}}$. For systems containing a very massive black hole, i.e., ${s_1=1/2,}~{s'_1=s''_1=0}$ and $m_1\gg m_2$, we recover the phase up to 2PN relative order derived in Ref.~\cite{Yunes2011}.

The computation of the Fourier-domain amplitude closely follows that of the dipole-driven regime. In place of $\sqrt{\rho(v)}$, one instead uses
\be
	\sqrt{\rho(v)}\simeq\sqrt{\rho^\text{non-dip}(v)}\left[1+\frac{1}{2}\frac{\rho^\text{dip}(v)}{\rho^\text{non-dip}(v)}\right].
\ee
Finally, one re-expands this expression using Eqs.~\eqref{eq:structrhoxnondip} and~\eqref{eq:structrhoxdip} and inserts the result in to Eq.~\eqref{eq:SPAamp}.

%%%%%%%%%%%%%%%%%%%%%%%%%%%%%%%%%%%%

\section{Conclusions}\label{sec:conclusions}

We have computed the gravitational waveform at 2PN relative order for a compact binary system on quasi-circular orbits in scalar-tensor theories with a single massless scalar. The phase and amplitude are presented in ready-to-use form for all $h_{lm}$ modes. We used the stationary phase approximation to express the waveform in Fourier space. We performed these calculations for systems whose inspiral is driven by the emission of dipolar radiation and those driven by quadrupolar flux. Because of the tight constraints on scalar-tensor gravity, only very low-frequency systems (e.g., binary pulsars) or those that host non-perturbative scalarization phenomena (e.g., spontaneous or dynamical scalarization) fall within this first regime --- most prospective GW sources will be quadrupolar-driven.

We conclude with a brief discussion of the potential utility of our results for testing GR with GWs. The early inspiral offers the best prospects for detecting the emission of dipolar radiation by compact binary systems, as radiation reaction enters at lower PN order in scalar-tensor theories than in GR. Thus, the best constraints would come from observation of neutron star-black hole binaries with space-based detectors.\footnote{Up to 2PN order, the signal produced by binary black holes is known to be identical to that in GR up to an undetectable rescaling of the gravitational constant $G$. \cite{Lang2014a}} Current estimates on the detectability of scalar-tensor effects have predominantly been made using the leading-order correction to the GW phase \cite{Will1994,Scharre2001,will2004, Berti2004, yagi2009} (although Ref.~\cite{Yunes2011} used 3.5PN scalar-tensor waveforms in studying extreme mass ratio inspirals).

Using Eq.~\eqref{eq:Psilmtotal}, we estimate the upper bound on the contribution of each PN correction to the phase. We consider the phase accumulated by a $100-1.4 \Msol$ system during an observation period of one year, spanning the frequency range $f\in(0.065 \text{Hz},1\text{Hz})$, subject to the experimental constraints discussed in Sec.~\ref{sec:constraints}. Relative to the $7.7\times 10^{6}$ cycles produced by the Newtonian-order GR term, the leading-order scalar-tensor correction decreases the phase by up to $\sim600$ GW cycles.\footnote{Note that this bound comes from assuming a neutron-star scalar charge of $\alpha_A\sim6\times 10^{-3}$, whereas Refs.~\cite{Will1994,Scharre2001,will2004, Berti2004, yagi2009} considered the maximum charge found in Brans-Dicke gravity $\alpha_A\sim 2\times10^{-3}$.} The 1PN relative order correction increases the total phase by another $\sim2$ cycles, although this piece would be difficult to detect, as it takes the same form as the leading-order GR term. The 1.5PN relative order correction adds $\sim 3$ GW cycles to the inspiral, and the 2PN order effect is below the limit of eLISA detectability, only contributing $\sim0.1$ cycles over the year. We emphasize that these values are only an order-of-magnitude estimate of the possible impact of higher-order scalar-tensor corrections; a more extensive parameter estimation study is needed to truly determine the detectability of these effects.

%%%%%%%%%%%%%%%%%%%%%%%%%%%%%%%%%%%%

\section{Acknowledgements}
\label{sec:acknowledgements}

%This work relies heavily on the foundation set by Ryan Lang, Saeed Mirshekari, and Clifford Will in Refs.~\cite{Mirshekari2013,Lang2014a,Lang2015a}; we wish to acknowledge their large-scale effort that made our results possible.

We are grateful to Ryan Lang for providing a \emph{Mathematica} notebook containing the results of Ref. \cite{Lang2015a} and to Lijing Shao for useful discussions concerning current binary pulsar constraints. N.S. acknowledges support from NSF Grant No. PHY-1208881.  S.M. acknowledges support from 
NASA grant 11-ATP-046 and NASA grant NNX12AN10G at the University of Maryland, College Park. N.S. thanks the Max Planck Institute for Gravitational Physics for its hospitality during the completion of this work.

%%%%%%%%%%%%%%%%%%%%%%%%%%%%%%%%%%%%

\appendix

\section{Translating notation}\label{app:notation}
This appendix contains the conversion between notation introduced in Table \ref{table:parameters} and that employed by Damour and Esposito-Far\`{e}se \cite{Damour1992,Damour1996}. Note that these authors defined $\alpha_A$ with a relative minus sign compared to Eq.~\eqref{eq:Chargedef} and defined $G_*$ as the bare gravitational constant in the Einstein frame.

\noindent\emph{Weak-field parameters}:
\begin{align}
G&\rightarrow \tilde{G}=G_* A_0^2\left(1+\alpha _0^2\right),\\
\zeta&\rightarrow\frac{\alpha _0^2}{1+\alpha _0^2},\\
\lambda_1&\rightarrow-\frac{\alpha _0 \beta _0}{\left(1+\alpha _0^2\right) A_0^2},\\
\lambda_2&\rightarrow-\frac{\alpha _0^2 \left(\alpha _0 \beta '_0-3 \beta _0^2\right)}{\left(1+\alpha _0^2\right)^2 A_0^4}.
\end{align}

\emph{Strong-field parameters}:
\begin{align}
s_A&\rightarrow\frac{1}{2}-\frac{\alpha _A}{2 \alpha _0},\\
s'_A&\rightarrow\frac{\beta _0 \alpha _A}{2 \alpha _0^2 A_0^2}+\frac{\beta _A}{4 \alpha _0^2},\\
s''_A&\rightarrow\frac{\alpha _A \beta '_0}{2 \alpha _0^2 A_0^4}+\frac{\beta '_A}{8 \alpha _0^3}-\frac{\beta _0^2 \alpha
   _A}{\alpha _0^3 A_0^4}+\frac{\beta _0 \alpha _A}{2 \alpha _0^2 A_0^2}-\frac{3 \beta _0 \beta _A}{4 \alpha
   _0^3 A_0^2}.
\end{align}

\emph{Binary parameters}:
\begin{align}
\alpha &\rightarrow\frac{1+\alpha_1\alpha_2}{1+\alpha_0^2},\\
{\gamma} &\rightarrow\gamma_{12}=-\frac{2 \alpha_1 \alpha_2}{1+\alpha_1\alpha_2},\\
{\beta}_1&\rightarrow\beta^{1}_{22}=\frac{\beta_1 \alpha_2^2}{2(1+\alpha_1 \alpha_2)^2},\\
{\beta}_2&\rightarrow\beta^{2}_{11}=\frac{\beta_2 \alpha_1^2}{2(1+\alpha_1 \alpha_2)^2},\\
{\delta}_1&\rightarrow\frac{\alpha_1^2}{(1+\alpha_1\alpha_2)^2},\\
{\delta}_2&\rightarrow\frac{\alpha_2^2}{(1+\alpha_1\alpha_2)^2},\\
{\chi}_1&\rightarrow-\frac{1}{4}\epsilon^{1}_{222}=-\frac{\beta'_1 \alpha_2^3}{4(1+\alpha_1 \alpha_2)^3},\\
{\chi}_2&\rightarrow-\frac{1}{4}\epsilon^{2}_{111}=-\frac{\beta'_2 \alpha_1^3}{4(1+\alpha_1 \alpha_2)^3}.
\end{align}
Our conversions agree with those presented in Table II of Ref. \cite{Mirshekari2013} with one exception: we find $G\alpha~=~\tilde{G}_{12}$ rather than  $G\alpha=G_{12}$.
%%%%%%%%%%%%%%%%%%%%%%%%%%%%%%%%%%%%

\begin{widetext}

\section{Explicit formulas for the Fourier-domain phasing}\label{app:coeffs}

This appendix gathers explicit formulae for the coefficients entering our results that were too voluminous to be kept in the main text.

In the dipole-driven case of Section~\ref{subsec:phasingDD}, we obtained for the flux
\begin{equation}
  \calF^{\rm DD} (x) = \frac{4  \mathcal{S}_-^2 \zeta \eta^2 x^4}{3 G \alpha} \left[ 1  + f^{\rm DD}_{2} x^{} + f^{\rm DD}_{3} x^{3/2} + f^{\rm DD}_{4} x^{2} + \mathcal{O}(5)\right]
\end{equation}

with the coefficients
\begin{subequations}\label{eq:coeffsfluxDD}
\begin{align}
f^{\rm DD}_{2}={}&- \frac{14}{5}
 + \frac{4 \mathcal{S}_+^2}{5 \mathcal{S}_-^2}
 -  \frac{4}{3} \beta_+
 + \frac{4 \mathcal{S}_+ \beta_-}{\mathcal{S}_- \gamma}
 + \frac{4 \beta_+}{\gamma}
 -  \frac{2}{3} \gamma
 + \frac{24}{5 \mathcal{S}_-^2 \zeta}
 + \frac{12 \gamma}{5 \mathcal{S}_-^2 \zeta}
 -  \frac{4}{3} \eta
 + \frac{4}{3} \beta_- \psi
 -  \frac{4 \beta_- \psi}{\gamma}
 -  \frac{4 \mathcal{S}_+ \beta_+ \psi}{\mathcal{S}_- \gamma},\\
f^{\rm DD}_{3}={}&2 \pi
 + \pi \gamma,\\
f^{\rm DD}_{4}={}&- \frac{29}{28}
 -  \frac{97 \mathcal{S}_+^2}{28 \mathcal{S}_-^2}
 -  \frac{4 \mathcal{S}_+ \beta_-}{3 \mathcal{S}_-}
 -  \frac{4}{3} \beta_-^2
 + \frac{2}{15} \beta_+
 -  \frac{32 \mathcal{S}_+^2 \beta_+}{15 \mathcal{S}_-^2}
 -  \frac{4}{3} \beta_+^2
 + \frac{4 \beta_-^2}{\gamma^2}
 + \frac{4 \mathcal{S}_+^2 \beta_-^2}{\mathcal{S}_-^2 \gamma^2}
 + \frac{16 \mathcal{S}_+ \beta_- \beta_+}{\mathcal{S}_- \gamma^2}
 + \frac{4 \beta_+^2}{\gamma^2}\nonumber\\
& + \frac{4 \mathcal{S}_+^2 \beta_+^2}{\mathcal{S}_-^2 \gamma^2}
 -  \frac{4 \mathcal{S}_+ \beta_-}{\mathcal{S}_- \gamma}
 -  \frac{8 \beta_-^2}{3 \gamma}
 -  \frac{36 \beta_+}{5 \gamma}
 + \frac{16 \mathcal{S}_+^2 \beta_+}{5 \mathcal{S}_-^2 \gamma}
 -  \frac{16 \mathcal{S}_+ \beta_- \beta_+}{3 \mathcal{S}_- \gamma}
 -  \frac{8 \beta_+^2}{3 \gamma}
 + \frac{2}{5} \gamma
 -  \frac{16 \mathcal{S}_+^2 \gamma}{15 \mathcal{S}_-^2}
 + \frac{1}{2} \gamma^2
 + \frac{2}{3} \delta_+\nonumber\\
& -  \frac{1247}{70 \mathcal{S}_-^2 \zeta}
 -  \frac{64 \beta_+}{5 \mathcal{S}_-^2 \zeta}
 -  \frac{2143 \gamma}{140 \mathcal{S}_-^2 \zeta}
 -  \frac{32 \beta_+ \gamma}{5 \mathcal{S}_-^2 \zeta}
 -  \frac{16 \gamma^2}{5 \mathcal{S}_-^2 \zeta}
 + \frac{55}{6} \eta
 -  \frac{7 \mathcal{S}_+^2 \eta}{3 \mathcal{S}_-^2}
 + \frac{16}{3} \beta_-^2 \eta
 + \frac{40}{3} \beta_+ \eta
 -  \frac{48 \beta_-^2 \eta}{\gamma^2}\nonumber\\
& -  \frac{32 \mathcal{S}_+ \beta_- \beta_+ \eta}{\mathcal{S}_- \gamma^2}
 + \frac{32 \beta_+^2 \eta}{\gamma^2}
 -  \frac{16 \mathcal{S}_+^2 \beta_+^2 \eta}{\mathcal{S}_-^2 \gamma^2}
 -  \frac{56 \mathcal{S}_+ \beta_- \eta}{3 \mathcal{S}_- \gamma}
 + \frac{80 \beta_-^2 \eta}{3 \gamma}
 -  \frac{56 \beta_+ \eta}{3 \gamma}
 + \frac{32 \mathcal{S}_+ \beta_- \beta_+ \eta}{3 \mathcal{S}_- \gamma}
 -  \frac{16 \beta_+^2 \eta}{\gamma}
 -  \frac{4}{3} \gamma \eta\nonumber\\
& -  \frac{1}{3} \gamma^2 \eta
 -  \frac{4}{3} \delta_+ \eta
 -  \frac{14 \eta}{\mathcal{S}_-^2 \zeta}
 -  \frac{7 \gamma \eta}{\mathcal{S}_-^2 \zeta}
 + \frac{2}{3} \eta^2
 + \frac{4 \mathcal{S}_+ \chi_-}{\mathcal{S}_- \gamma}
 -  \frac{8 \mathcal{S}_+ \eta \chi_-}{\mathcal{S}_- \gamma}
 -  \frac{4}{3} \chi_+
 + \frac{4 \chi_+}{\gamma}
 + \frac{8}{3} \eta \chi_+
 -  \frac{8 \eta \chi_+}{\gamma}
 + \frac{11 \mathcal{S}_+ \psi}{2 \mathcal{S}_-}\nonumber\\
& -  \frac{2}{15} \beta_- \psi
 + \frac{32 \mathcal{S}_+^2 \beta_- \psi}{15 \mathcal{S}_-^2}
 + \frac{4 \mathcal{S}_+ \beta_+ \psi}{3 \mathcal{S}_-}
 + \frac{8}{3} \beta_- \beta_+ \psi
 -  \frac{8 \mathcal{S}_+ \beta_-^2 \psi}{\mathcal{S}_- \gamma^2}
 -  \frac{8 \beta_- \beta_+ \psi}{\gamma^2}
 -  \frac{8 \mathcal{S}_+^2 \beta_- \beta_+ \psi}{\mathcal{S}_-^2 \gamma^2}
 -  \frac{8 \mathcal{S}_+ \beta_+^2 \psi}{\mathcal{S}_- \gamma^2}\nonumber\\
& + \frac{36 \beta_- \psi}{5 \gamma}
 -  \frac{16 \mathcal{S}_+^2 \beta_- \psi}{5 \mathcal{S}_-^2 \gamma}
 + \frac{8 \mathcal{S}_+ \beta_-^2 \psi}{3 \mathcal{S}_- \gamma}
 + \frac{4 \mathcal{S}_+ \beta_+ \psi}{\mathcal{S}_- \gamma}
 + \frac{16 \beta_- \beta_+ \psi}{3 \gamma}
 + \frac{8 \mathcal{S}_+ \beta_+^2 \psi}{3 \mathcal{S}_- \gamma}
 + \frac{3 \mathcal{S}_+ \gamma \psi}{\mathcal{S}_-}
 + \frac{2}{3} \delta_- \psi
 + \frac{64 \beta_- \psi}{5 \mathcal{S}_-^2 \zeta}\nonumber\\
& + \frac{32 \beta_- \gamma \psi}{5 \mathcal{S}_-^2 \zeta}
 -  \frac{16}{3} \beta_- \eta \psi
 + \frac{20 \beta_- \eta \psi}{3 \gamma}
 + \frac{20 \mathcal{S}_+ \beta_+ \eta \psi}{3 \mathcal{S}_- \gamma}
 + \frac{4}{3} \chi_- \psi
 -  \frac{4 \chi_- \psi}{\gamma}
 -  \frac{4 \mathcal{S}_+ \chi_+ \psi}{\mathcal{S}_- \gamma}.
\end{align}
\end{subequations}

For the ratio $\rho(x)$, we obtained
\begin{equation}
  \rho^{\rm DD} (x) = \frac{3}{8 \mathcal{S}_-^2 \zeta \eta x^4} \left[ 1  + \rho^{\rm DD}_{2} x^{} + \rho^{\rm DD}_{3} x^{3/2} + \rho^{\rm DD}_{4} x^{2} + \mathcal{O}(5)\right]
\end{equation}

with the coefficients
\begin{subequations}\label{eq:coeffsrhoDD}
\begin{align}
\rho^{\rm DD}_{2}={}&\frac{13}{10}
 -  \frac{4 \mathcal{S}_+^2}{5 \mathcal{S}_-^2}
 + \frac{8}{3} \beta_+
 -  \frac{4 \mathcal{S}_+ \beta_-}{\mathcal{S}_- \gamma}
 -  \frac{4 \beta_+}{\gamma}
 -  \frac{2}{3} \gamma
 -  \frac{24}{5 \mathcal{S}_-^2 \zeta}
 -  \frac{12 \gamma}{5 \mathcal{S}_-^2 \zeta}
 + \frac{7}{6} \eta
 -  \frac{8}{3} \beta_- \psi
 + \frac{4 \beta_- \psi}{\gamma}
 + \frac{4 \mathcal{S}_+ \beta_+ \psi}{\mathcal{S}_- \gamma},\\
\rho^{\rm DD}_{3}={}&-2 \pi
 -  \pi \gamma,\\
\rho^{\rm DD}_{4}={}&- \frac{7629}{1400}
 + \frac{129 \mathcal{S}_+^2}{700 \mathcal{S}_-^2}
 + \frac{16 \mathcal{S}_+^4}{25 \mathcal{S}_-^4}
 + \frac{4 \mathcal{S}_+ \beta_-}{3 \mathcal{S}_-}
 + \frac{80}{9} \beta_-^2
 + \frac{181}{15} \beta_+
 -  \frac{16 \mathcal{S}_+^2 \beta_+}{15 \mathcal{S}_-^2}
 + \frac{80}{9} \beta_+^2
 + \frac{12 \beta_-^2}{\gamma^2}
 + \frac{12 \mathcal{S}_+^2 \beta_-^2}{\mathcal{S}_-^2 \gamma^2}\nonumber\\
& + \frac{48 \mathcal{S}_+ \beta_- \beta_+}{\mathcal{S}_- \gamma^2}
 + \frac{12 \beta_+^2}{\gamma^2}
 + \frac{12 \mathcal{S}_+^2 \beta_+^2}{\mathcal{S}_-^2 \gamma^2}
 -  \frac{62 \mathcal{S}_+ \beta_-}{5 \mathcal{S}_- \gamma}
 + \frac{32 \mathcal{S}_+^3 \beta_-}{5 \mathcal{S}_-^3 \gamma}
 -  \frac{40 \beta_-^2}{3 \gamma}
 -  \frac{46 \beta_+}{5 \gamma}
 + \frac{16 \mathcal{S}_+^2 \beta_+}{5 \mathcal{S}_-^2 \gamma}
 -  \frac{80 \mathcal{S}_+ \beta_- \beta_+}{3 \mathcal{S}_- \gamma}\nonumber\\
& -  \frac{40 \beta_+^2}{3 \gamma}
 -  \frac{77}{5} \gamma
 + \frac{16 \mathcal{S}_+^2 \gamma}{15 \mathcal{S}_-^2}
 + \frac{44}{9} \beta_+ \gamma
 -  \frac{205}{36} \gamma^2
 + \frac{1}{3} \delta_+
 + \frac{576}{25 \mathcal{S}_-^4 \zeta^2}
 + \frac{576 \gamma}{25 \mathcal{S}_-^4 \zeta^2}
 + \frac{144 \gamma^2}{25 \mathcal{S}_-^4 \zeta^2}
 -  \frac{653}{350 \mathcal{S}_-^2 \zeta}
 + \frac{192 \mathcal{S}_+^2}{25 \mathcal{S}_-^4 \zeta}\nonumber\\
& + \frac{96 \mathcal{S}_+ \beta_-}{5 \mathcal{S}_-^3 \zeta}
 + \frac{64 \beta_+}{5 \mathcal{S}_-^2 \zeta}
 + \frac{192 \mathcal{S}_+ \beta_-}{5 \mathcal{S}_-^3 \gamma \zeta}
 + \frac{192 \beta_+}{5 \mathcal{S}_-^2 \gamma \zeta}
 + \frac{3827 \gamma}{700 \mathcal{S}_-^2 \zeta}
 + \frac{96 \mathcal{S}_+^2 \gamma}{25 \mathcal{S}_-^4 \zeta}
 -  \frac{16 \beta_+ \gamma}{5 \mathcal{S}_-^2 \zeta}
 + \frac{16 \gamma^2}{5 \mathcal{S}_-^2 \zeta}
 + \frac{71}{24} \eta
 + \frac{\mathcal{S}_+^2 \eta}{3 \mathcal{S}_-^2}\nonumber\\
& -  \frac{320}{9} \beta_-^2 \eta
 -  \frac{245}{9} \beta_+ \eta
 -  \frac{16 \beta_-^2 \eta}{\gamma^2}
 -  \frac{96 \mathcal{S}_+ \beta_- \beta_+ \eta}{\mathcal{S}_- \gamma^2}
 -  \frac{32 \beta_+^2 \eta}{\gamma^2}
 -  \frac{48 \mathcal{S}_+^2 \beta_+^2 \eta}{\mathcal{S}_-^2 \gamma^2}
 + \frac{26 \mathcal{S}_+ \beta_- \eta}{3 \mathcal{S}_- \gamma}
 -  \frac{32 \beta_-^2 \eta}{3 \gamma}
 + \frac{26 \beta_+ \eta}{3 \gamma}\nonumber\\
& + \frac{160 \mathcal{S}_+ \beta_- \beta_+ \eta}{3 \mathcal{S}_- \gamma}
 + \frac{64 \beta_+^2 \eta}{\gamma}
 + \frac{110}{9} \gamma \eta
 + \frac{4}{3} \gamma^2 \eta
 + \frac{16}{3} \delta_+ \eta
 + \frac{2 \eta}{\mathcal{S}_-^2 \zeta}
 + \frac{\gamma \eta}{\mathcal{S}_-^2 \zeta}
 + \frac{55}{72} \eta^2
 -  \frac{4 \mathcal{S}_+ \chi_-}{\mathcal{S}_- \gamma}
 + \frac{8 \mathcal{S}_+ \eta \chi_-}{\mathcal{S}_- \gamma}\nonumber\\
& + \frac{16}{3} \chi_+
 -  \frac{4 \chi_+}{\gamma}
 -  \frac{32}{3} \eta \chi_+
 + \frac{8 \eta \chi_+}{\gamma}
 -  \frac{11 \mathcal{S}_+ \psi}{2 \mathcal{S}_-}
 -  \frac{181}{15} \beta_- \psi
 + \frac{16 \mathcal{S}_+^2 \beta_- \psi}{15 \mathcal{S}_-^2}
 -  \frac{4 \mathcal{S}_+ \beta_+ \psi}{3 \mathcal{S}_-}
 -  \frac{160}{9} \beta_- \beta_+ \psi\nonumber\\
& -  \frac{24 \mathcal{S}_+ \beta_-^2 \psi}{\mathcal{S}_- \gamma^2}
 -  \frac{24 \beta_- \beta_+ \psi}{\gamma^2}
 -  \frac{24 \mathcal{S}_+^2 \beta_- \beta_+ \psi}{\mathcal{S}_-^2 \gamma^2}
 -  \frac{24 \mathcal{S}_+ \beta_+^2 \psi}{\mathcal{S}_- \gamma^2}
 + \frac{46 \beta_- \psi}{5 \gamma}
 -  \frac{16 \mathcal{S}_+^2 \beta_- \psi}{5 \mathcal{S}_-^2 \gamma}
 + \frac{40 \mathcal{S}_+ \beta_-^2 \psi}{3 \mathcal{S}_- \gamma}\nonumber\\
& + \frac{62 \mathcal{S}_+ \beta_+ \psi}{5 \mathcal{S}_- \gamma}
 -  \frac{32 \mathcal{S}_+^3 \beta_+ \psi}{5 \mathcal{S}_-^3 \gamma}
 + \frac{80 \beta_- \beta_+ \psi}{3 \gamma}
 + \frac{40 \mathcal{S}_+ \beta_+^2 \psi}{3 \mathcal{S}_- \gamma}
 -  \frac{3 \mathcal{S}_+ \gamma \psi}{\mathcal{S}_-}
 -  \frac{44}{9} \beta_- \gamma \psi
 + \frac{1}{3} \delta_- \psi
 -  \frac{64 \beta_- \psi}{5 \mathcal{S}_-^2 \zeta}\nonumber\\
& -  \frac{96 \mathcal{S}_+ \beta_+ \psi}{5 \mathcal{S}_-^3 \zeta}
 -  \frac{192 \beta_- \psi}{5 \mathcal{S}_-^2 \gamma \zeta}
 -  \frac{192 \mathcal{S}_+ \beta_+ \psi}{5 \mathcal{S}_-^3 \gamma \zeta}
 + \frac{16 \beta_- \gamma \psi}{5 \mathcal{S}_-^2 \zeta}
 + \frac{11}{9} \beta_- \eta \psi
 + \frac{10 \beta_- \eta \psi}{3 \gamma}
 + \frac{10 \mathcal{S}_+ \beta_+ \eta \psi}{3 \mathcal{S}_- \gamma}
 -  \frac{16}{3} \chi_- \psi\nonumber\\
& + \frac{4 \chi_- \psi}{\gamma}
 + \frac{4 \mathcal{S}_+ \chi_+ \psi}{\mathcal{S}_- \gamma}.
\end{align}
\end{subequations}

In the quadrupole-driven case of Section~\ref{subsec:phasingQD}, we obtained for the dipolar part of the flux
\begin{equation}
  \calF_\text{dip} (x) = \frac{4 \mathcal{S}_-^2 \zeta \eta^2 x^4}{3 G \alpha} \left[ 1  + f^{\rm d}_{2} x^{} + f^{\rm d}_{3} x^{3/2} + f^{\rm d}_{4} x^{2} + \mathcal{O}(5)\right]
\end{equation}

with coefficients given by
\begin{subequations}\label{eq:coeffsfluxdip}
\begin{align}
f^{\rm d}_{2}={}&- \frac{14}{5}
 -  \frac{4}{3} \beta_+
 + \frac{4 \mathcal{S}_+ \beta_-}{\mathcal{S}_- \gamma}
 + \frac{4 \beta_+}{\gamma}
 -  \frac{2}{3} \gamma
 -  \frac{4}{3} \eta
 + \frac{4}{3} \beta_- \psi
 -  \frac{4 \beta_- \psi}{\gamma}
 -  \frac{4 \mathcal{S}_+ \beta_+ \psi}{\mathcal{S}_- \gamma},\\
f^{\rm d}_{3}={}&2 \pi
 + \pi \gamma,\\
f^{\rm d}_{4}={}&- \frac{29}{28}
 -  \frac{4 \mathcal{S}_+ \beta_-}{3 \mathcal{S}_-}
 -  \frac{4}{3} \beta_-^2
 + \frac{2}{15} \beta_+
 -  \frac{4}{3} \beta_+^2
 + \frac{4 \beta_-^2}{\gamma^2}
 + \frac{16 \mathcal{S}_+ \beta_- \beta_+}{\mathcal{S}_- \gamma^2}
 + \frac{4 \beta_+^2}{\gamma^2}
 -  \frac{4 \mathcal{S}_+ \beta_-}{\mathcal{S}_- \gamma}
 -  \frac{8 \beta_-^2}{3 \gamma}
 -  \frac{36 \beta_+}{5 \gamma}\nonumber\\
& -  \frac{16 \mathcal{S}_+ \beta_- \beta_+}{3 \mathcal{S}_- \gamma}
 -  \frac{8 \beta_+^2}{3 \gamma}
 + \frac{2}{5} \gamma
 + \frac{1}{2} \gamma^2
 + \frac{2}{3} \delta_+
 + \frac{55}{6} \eta
 + \frac{16}{3} \beta_-^2 \eta
 + \frac{40}{3} \beta_+ \eta
 -  \frac{48 \beta_-^2 \eta}{\gamma^2}
 -  \frac{32 \mathcal{S}_+ \beta_- \beta_+ \eta}{\mathcal{S}_- \gamma^2}
 + \frac{32 \beta_+^2 \eta}{\gamma^2}\nonumber\\
& -  \frac{56 \mathcal{S}_+ \beta_- \eta}{3 \mathcal{S}_- \gamma}
 + \frac{80 \beta_-^2 \eta}{3 \gamma}
 -  \frac{56 \beta_+ \eta}{3 \gamma}
 + \frac{32 \mathcal{S}_+ \beta_- \beta_+ \eta}{3 \mathcal{S}_- \gamma}
 -  \frac{16 \beta_+^2 \eta}{\gamma}
 -  \frac{4}{3} \gamma \eta
 -  \frac{1}{3} \gamma^2 \eta
 -  \frac{4}{3} \delta_+ \eta
 + \frac{2}{3} \eta^2
 + \frac{4 \mathcal{S}_+ \chi_-}{\mathcal{S}_- \gamma}\nonumber\\
& -  \frac{8 \mathcal{S}_+ \eta \chi_-}{\mathcal{S}_- \gamma}
 -  \frac{4}{3} \chi_+
 + \frac{4 \chi_+}{\gamma}
 + \frac{8}{3} \eta \chi_+
 -  \frac{8 \eta \chi_+}{\gamma}
 + \frac{11 \mathcal{S}_+ \psi}{2 \mathcal{S}_-}
 -  \frac{2}{15} \beta_- \psi
 + \frac{4 \mathcal{S}_+ \beta_+ \psi}{3 \mathcal{S}_-}
 + \frac{8}{3} \beta_- \beta_+ \psi
 -  \frac{8 \mathcal{S}_+ \beta_-^2 \psi}{\mathcal{S}_- \gamma^2}\nonumber\\
& -  \frac{8 \beta_- \beta_+ \psi}{\gamma^2}
 -  \frac{8 \mathcal{S}_+ \beta_+^2 \psi}{\mathcal{S}_- \gamma^2}
 + \frac{36 \beta_- \psi}{5 \gamma}
 + \frac{8 \mathcal{S}_+ \beta_-^2 \psi}{3 \mathcal{S}_- \gamma}
 + \frac{4 \mathcal{S}_+ \beta_+ \psi}{\mathcal{S}_- \gamma}
 + \frac{16 \beta_- \beta_+ \psi}{3 \gamma}
 + \frac{8 \mathcal{S}_+ \beta_+^2 \psi}{3 \mathcal{S}_- \gamma}
 + \frac{3 \mathcal{S}_+ \gamma \psi}{\mathcal{S}_-}
 + \frac{2}{3} \delta_- \psi\nonumber\\
& -  \frac{16}{3} \beta_- \eta \psi
 + \frac{20 \beta_- \eta \psi}{3 \gamma}
 + \frac{20 \mathcal{S}_+ \beta_+ \eta \psi}{3 \mathcal{S}_- \gamma}
 + \frac{4}{3} \chi_- \psi
 -  \frac{4 \chi_- \psi}{\gamma}
 -  \frac{4 \mathcal{S}_+ \chi_+ \psi}{\mathcal{S}_- \gamma}.
\end{align}
\end{subequations}

For the non-dipolar part, we obtained
\begin{equation}
  \calF_\text{non-dip} (x) = \frac{32  \eta^2 \xi x^5}{5 G \alpha} \left[ 1  + f^{\rm nd}_{2} x^{} + \mathcal{O}(3)\right]
\end{equation}

with the coefficient
\begin{subequations}\label{eq:coeffsfluxnondip}
\begin{align}
f^{\rm nd}_{2}={}&- \frac{1247}{336 \xi}
 -  \frac{8 \beta_+}{3 \xi}
 -  \frac{2143 \gamma}{672 \xi}
 -  \frac{4 \beta_+ \gamma}{3 \xi}
 -  \frac{2 \gamma^2}{3 \xi}
 -  \frac{485 \mathcal{S}_+^2 \zeta}{672 \xi}
 -  \frac{4 \mathcal{S}_+^2 \beta_+ \zeta}{9 \xi}
 + \frac{5 \mathcal{S}_+^2 \beta_-^2 \zeta}{6 \gamma^2 \xi}
 + \frac{5 \mathcal{S}_+^2 \beta_+^2 \zeta}{6 \gamma^2 \xi}
 + \frac{2 \mathcal{S}_+^2 \beta_+ \zeta}{3 \gamma \xi}\nonumber\\
& -  \frac{2 \mathcal{S}_+^2 \gamma \zeta}{9 \xi}
 -  \frac{35 \eta}{12 \xi}
 -  \frac{35 \gamma \eta}{24 \xi}
 -  \frac{35 \mathcal{S}_+^2 \zeta \eta}{72 \xi}
 -  \frac{10 \mathcal{S}_+^2 \beta_+^2 \zeta \eta}{3 \gamma^2 \xi}
 + \frac{8 \beta_- \psi}{3 \xi}
 + \frac{4 \beta_- \gamma \psi}{3 \xi}
 + \frac{4 \mathcal{S}_+^2 \beta_- \zeta \psi}{9 \xi}
 -  \frac{5 \mathcal{S}_+^2 \beta_- \beta_+ \zeta \psi}{3 \gamma^2 \xi}\nonumber\\
& -  \frac{2 \mathcal{S}_+^2 \beta_- \zeta \psi}{3 \gamma \xi}.
\end{align}
\end{subequations}

For the non-dipolar ratio $\rho_\text{non-dip}(x)$, we obtain
\begin{equation}
  \rho_\text{non-dip} (x) = \frac{5}{64 \eta \xi x^5} \left[ 1  + \rho^{\rm nd}_{2} x^{} + \rho^{\rm nd}_{3} x^{3/2} + \rho^{\rm nd}_{4} x^{2} + \mathcal{O}(5)\right]
\end{equation}

with the coefficients
\begin{subequations}\label{eq:coeffsrhonondip}
\begin{align}
\rho^{\rm nd}_{2}={}&\frac{743}{336 \xi}
 + \frac{4 \beta_+}{\xi}
 + \frac{743 \gamma}{672 \xi}
 + \frac{2 \beta_+ \gamma}{\xi}
 + \frac{317 \mathcal{S}_+^2 \zeta}{672 \xi}
 + \frac{2 \mathcal{S}_+^2 \beta_+ \zeta}{3 \xi}
 -  \frac{5 \mathcal{S}_+^2 \beta_-^2 \zeta}{6 \gamma^2 \xi}
 -  \frac{5 \mathcal{S}_+^2 \beta_+^2 \zeta}{6 \gamma^2 \xi}
 -  \frac{2 \mathcal{S}_+^2 \beta_+ \zeta}{3 \gamma \xi}
 + \frac{11 \eta}{4 \xi}
 + \frac{11 \gamma \eta}{8 \xi}\nonumber\\
& + \frac{11 \mathcal{S}_+^2 \zeta \eta}{24 \xi}
 + \frac{10 \mathcal{S}_+^2 \beta_+^2 \zeta \eta}{3 \gamma^2 \xi}
 -  \frac{4 \beta_- \psi}{\xi}
 -  \frac{2 \beta_- \gamma \psi}{\xi}
 -  \frac{2 \mathcal{S}_+^2 \beta_- \zeta \psi}{3 \xi}
 + \frac{5 \mathcal{S}_+^2 \beta_- \beta_+ \zeta \psi}{3 \gamma^2 \xi}
 + \frac{2 \mathcal{S}_+^2 \beta_- \zeta \psi}{3 \gamma \xi},\\
\rho^{\rm nd}_{3}={}&- f^{\rm ST}_{3}
 -  \frac{4 \pi}{\xi},\\
\rho^{\rm nd}_{4}={}&- \frac{81}{8}
 -  f^{\rm ST}_{4}
 + 4 \beta_-^2
 + 3 \beta_+
 + 4 \beta_+^2
 - 14 \gamma
 + 4 \beta_+ \gamma
 -  \frac{19}{4} \gamma^2
 + \delta_+
 + \frac{57}{8} \eta
 - 16 \beta_-^2 \eta
 - 19 \beta_+ \eta
 -  \frac{48 \beta_-^2 \eta}{\gamma}\nonumber\\
& + \frac{48 \beta_+^2 \eta}{\gamma}
 + 11 \gamma \eta
 + \gamma^2 \eta
 + 4 \delta_+ \eta
 -  \frac{1}{8} \eta^2
 + \frac{1555009}{112896 \xi^2}
 + \frac{64 \beta_-^2}{9 \xi^2}
 + \frac{1247 \beta_+}{63 \xi^2}
 + \frac{64 \beta_+^2}{9 \xi^2}
 + \frac{2672321 \gamma}{112896 \xi^2}
 + \frac{64 \beta_-^2 \gamma}{9 \xi^2}\nonumber\\
& + \frac{565 \beta_+ \gamma}{21 \xi^2}
 + \frac{64 \beta_+^2 \gamma}{9 \xi^2}
 + \frac{2275691 \gamma^2}{150528 \xi^2}
 + \frac{16 \beta_-^2 \gamma^2}{9 \xi^2}
 + \frac{1013 \beta_+ \gamma^2}{84 \xi^2}
 + \frac{16 \beta_+^2 \gamma^2}{9 \xi^2}
 + \frac{2143 \gamma^3}{504 \xi^2}
 + \frac{16 \beta_+ \gamma^3}{9 \xi^2}
 + \frac{4 \gamma^4}{9 \xi^2}\nonumber\\
& + \frac{604795 \mathcal{S}_+^2 \zeta}{112896 \xi^2}
 -  \frac{14 \mathcal{S}_+^2 \beta_-^2 \zeta}{27 \xi^2}
 + \frac{4379 \mathcal{S}_+^2 \beta_+ \zeta}{1512 \xi^2}
 -  \frac{14 \mathcal{S}_+^2 \beta_+^2 \zeta}{27 \xi^2}
 -  \frac{6235 \mathcal{S}_+^2 \beta_-^2 \zeta}{1008 \gamma^2 \xi^2}
 -  \frac{40 \mathcal{S}_+^2 \beta_-^2 \beta_+ \zeta}{3 \gamma^2 \xi^2}
 -  \frac{6235 \mathcal{S}_+^2 \beta_+^2 \zeta}{1008 \gamma^2 \xi^2}\nonumber\\
& -  \frac{40 \mathcal{S}_+^2 \beta_+^3 \zeta}{9 \gamma^2 \xi^2}
 -  \frac{1987 \mathcal{S}_+^2 \beta_-^2 \zeta}{224 \gamma \xi^2}
 -  \frac{1247 \mathcal{S}_+^2 \beta_+ \zeta}{252 \gamma \xi^2}
 -  \frac{20 \mathcal{S}_+^2 \beta_-^2 \beta_+ \zeta}{3 \gamma \xi^2}
 -  \frac{1987 \mathcal{S}_+^2 \beta_+^2 \zeta}{224 \gamma \xi^2}
 -  \frac{20 \mathcal{S}_+^2 \beta_+^3 \zeta}{9 \gamma \xi^2}
 + \frac{4235377 \mathcal{S}_+^2 \gamma \zeta}{677376 \xi^2}\nonumber\\
& + \frac{32 \mathcal{S}_+^2 \beta_-^2 \gamma \zeta}{27 \xi^2}
 + \frac{91 \mathcal{S}_+^2 \beta_+ \gamma \zeta}{18 \xi^2}
 + \frac{32 \mathcal{S}_+^2 \beta_+^2 \gamma \zeta}{27 \xi^2}
 + \frac{257 \mathcal{S}_+^2 \gamma^2 \zeta}{108 \xi^2}
 + \frac{32 \mathcal{S}_+^2 \beta_+ \gamma^2 \zeta}{27 \xi^2}
 + \frac{8 \mathcal{S}_+^2 \gamma^3 \zeta}{27 \xi^2}
 + \frac{235225 \mathcal{S}_+^4 \zeta^2}{451584 \xi^2}\nonumber\\
& + \frac{16 \mathcal{S}_+^4 \beta_-^2 \zeta^2}{81 \xi^2}
 + \frac{29 \mathcal{S}_+^4 \beta_+ \zeta^2}{84 \xi^2}
 + \frac{16 \mathcal{S}_+^4 \beta_+^2 \zeta^2}{81 \xi^2}
 + \frac{25 \mathcal{S}_+^4 \beta_-^4 \zeta^2}{36 \gamma^4 \xi^2}
 + \frac{25 \mathcal{S}_+^4 \beta_-^2 \beta_+^2 \zeta^2}{6 \gamma^4 \xi^2}
 + \frac{25 \mathcal{S}_+^4 \beta_+^4 \zeta^2}{36 \gamma^4 \xi^2}
 + \frac{10 \mathcal{S}_+^4 \beta_-^2 \beta_+ \zeta^2}{3 \gamma^3 \xi^2}\nonumber\\
& + \frac{10 \mathcal{S}_+^4 \beta_+^3 \zeta^2}{9 \gamma^3 \xi^2}
 -  \frac{1529 \mathcal{S}_+^4 \beta_-^2 \zeta^2}{2016 \gamma^2 \xi^2}
 -  \frac{20 \mathcal{S}_+^4 \beta_-^2 \beta_+ \zeta^2}{9 \gamma^2 \xi^2}
 -  \frac{1529 \mathcal{S}_+^4 \beta_+^2 \zeta^2}{2016 \gamma^2 \xi^2}
 -  \frac{20 \mathcal{S}_+^4 \beta_+^3 \zeta^2}{27 \gamma^2 \xi^2}
 -  \frac{26 \mathcal{S}_+^4 \beta_-^2 \zeta^2}{27 \gamma \xi^2}\nonumber\\
& -  \frac{485 \mathcal{S}_+^4 \beta_+ \zeta^2}{504 \gamma \xi^2}
 -  \frac{26 \mathcal{S}_+^4 \beta_+^2 \zeta^2}{27 \gamma \xi^2}
 + \frac{485 \mathcal{S}_+^4 \gamma \zeta^2}{1512 \xi^2}
 + \frac{16 \mathcal{S}_+^4 \beta_+ \gamma \zeta^2}{81 \xi^2}
 + \frac{4 \mathcal{S}_+^4 \gamma^2 \zeta^2}{81 \xi^2}
 + \frac{6235 \eta}{288 \xi^2}
 -  \frac{256 \beta_-^2 \eta}{9 \xi^2}
 + \frac{140 \beta_+ \eta}{9 \xi^2}\nonumber\\
& + \frac{2825 \gamma \eta}{96 \xi^2}
 -  \frac{256 \beta_-^2 \gamma \eta}{9 \xi^2}
 + \frac{140 \beta_+ \gamma \eta}{9 \xi^2}
 + \frac{5065 \gamma^2 \eta}{384 \xi^2}
 -  \frac{64 \beta_-^2 \gamma^2 \eta}{9 \xi^2}
 + \frac{35 \beta_+ \gamma^2 \eta}{9 \xi^2}
 + \frac{35 \gamma^3 \eta}{18 \xi^2}
 + \frac{6755 \mathcal{S}_+^2 \zeta \eta}{864 \xi^2}\nonumber\\
& -  \frac{64 \mathcal{S}_+^2 \beta_-^2 \zeta \eta}{27 \xi^2}
 + \frac{175 \mathcal{S}_+^2 \beta_+ \zeta \eta}{54 \xi^2}
 + \frac{40 \mathcal{S}_+^2 \beta_+^2 \zeta \eta}{9 \xi^2}
 -  \frac{175 \mathcal{S}_+^2 \beta_-^2 \zeta \eta}{36 \gamma^2 \xi^2}
 + \frac{320 \mathcal{S}_+^2 \beta_-^2 \beta_+ \zeta \eta}{9 \gamma^2 \xi^2}
 + \frac{835 \mathcal{S}_+^2 \beta_+^2 \zeta \eta}{42 \gamma^2 \xi^2}\nonumber\\
& + \frac{160 \mathcal{S}_+^2 \beta_+^3 \zeta \eta}{9 \gamma^2 \xi^2}
 + \frac{283 \mathcal{S}_+^2 \beta_-^2 \zeta \eta}{24 \gamma \xi^2}
 -  \frac{35 \mathcal{S}_+^2 \beta_+ \zeta \eta}{9 \gamma \xi^2}
 + \frac{160 \mathcal{S}_+^2 \beta_-^2 \beta_+ \zeta \eta}{9 \gamma \xi^2}
 + \frac{4745 \mathcal{S}_+^2 \beta_+^2 \zeta \eta}{252 \gamma \xi^2}
 + \frac{80 \mathcal{S}_+^2 \beta_+^3 \zeta \eta}{9 \gamma \xi^2}\nonumber\\
& + \frac{3745 \mathcal{S}_+^2 \gamma \zeta \eta}{576 \xi^2}
 -  \frac{128 \mathcal{S}_+^2 \beta_-^2 \gamma \zeta \eta}{27 \xi^2}
 + \frac{70 \mathcal{S}_+^2 \beta_+ \gamma \zeta \eta}{27 \xi^2}
 + \frac{35 \mathcal{S}_+^2 \gamma^2 \zeta \eta}{27 \xi^2}
 + \frac{2425 \mathcal{S}_+^4 \zeta^2 \eta}{3456 \xi^2}
 -  \frac{64 \mathcal{S}_+^4 \beta_-^2 \zeta^2 \eta}{81 \xi^2}\nonumber\\
& + \frac{35 \mathcal{S}_+^4 \beta_+ \zeta^2 \eta}{81 \xi^2}
 -  \frac{50 \mathcal{S}_+^4 \beta_-^2 \beta_+^2 \zeta^2 \eta}{3 \gamma^4 \xi^2}
 -  \frac{50 \mathcal{S}_+^4 \beta_+^4 \zeta^2 \eta}{9 \gamma^4 \xi^2}
 -  \frac{80 \mathcal{S}_+^4 \beta_-^2 \beta_+ \zeta^2 \eta}{9 \gamma^3 \xi^2}
 -  \frac{40 \mathcal{S}_+^4 \beta_+^3 \zeta^2 \eta}{9 \gamma^3 \xi^2}
 -  \frac{559 \mathcal{S}_+^4 \beta_-^2 \zeta^2 \eta}{216 \gamma^2 \xi^2}\nonumber\\
& + \frac{160 \mathcal{S}_+^4 \beta_-^2 \beta_+ \zeta^2 \eta}{27 \gamma^2 \xi^2}
 + \frac{3025 \mathcal{S}_+^4 \beta_+^2 \zeta^2 \eta}{756 \gamma^2 \xi^2}
 + \frac{80 \mathcal{S}_+^4 \beta_+^3 \zeta^2 \eta}{27 \gamma^2 \xi^2}
 + \frac{64 \mathcal{S}_+^4 \beta_-^2 \zeta^2 \eta}{27 \gamma \xi^2}
 -  \frac{35 \mathcal{S}_+^4 \beta_+ \zeta^2 \eta}{54 \gamma \xi^2}
 + \frac{40 \mathcal{S}_+^4 \beta_+^2 \zeta^2 \eta}{27 \gamma \xi^2}\nonumber\\
& + \frac{35 \mathcal{S}_+^4 \gamma \zeta^2 \eta}{162 \xi^2}
 + \frac{1225 \eta^2}{144 \xi^2}
 + \frac{1225 \gamma \eta^2}{144 \xi^2}
 + \frac{1225 \gamma^2 \eta^2}{576 \xi^2}
 + \frac{1225 \mathcal{S}_+^2 \zeta \eta^2}{432 \xi^2}
 + \frac{175 \mathcal{S}_+^2 \beta_+^2 \zeta \eta^2}{9 \gamma^2 \xi^2}
 + \frac{175 \mathcal{S}_+^2 \beta_+^2 \zeta \eta^2}{18 \gamma \xi^2}\nonumber\\
& + \frac{1225 \mathcal{S}_+^2 \gamma \zeta \eta^2}{864 \xi^2}
 + \frac{1225 \mathcal{S}_+^4 \zeta^2 \eta^2}{5184 \xi^2}
 + \frac{100 \mathcal{S}_+^4 \beta_+^4 \zeta^2 \eta^2}{9 \gamma^4 \xi^2}
 + \frac{175 \mathcal{S}_+^4 \beta_+^2 \zeta^2 \eta^2}{54 \gamma^2 \xi^2}
 -  \frac{1655}{2592 \xi}
 + \frac{32 \beta_-^2}{9 \xi}
 + \frac{239 \beta_+}{252 \xi}
 + \frac{32 \beta_+^2}{9 \xi}\nonumber\\
& -  \frac{39239 \gamma}{4032 \xi}
 + \frac{16 \beta_-^2 \gamma}{9 \xi}
 -  \frac{73 \beta_+ \gamma}{56 \xi}
 + \frac{16 \beta_+^2 \gamma}{9 \xi}
 -  \frac{2647 \gamma^2}{504 \xi}
 -  \frac{8 \beta_+ \gamma^2}{9 \xi}
 -  \frac{8 \gamma^3}{9 \xi}
 -  \frac{485 \mathcal{S}_+^2 \zeta}{448 \xi}
 + \frac{16 \mathcal{S}_+^2 \beta_-^2 \zeta}{27 \xi}\nonumber\\
& + \frac{199 \mathcal{S}_+^2 \beta_+ \zeta}{168 \xi}
 + \frac{16 \mathcal{S}_+^2 \beta_+^2 \zeta}{27 \xi}
 + \frac{5 \mathcal{S}_+^2 \beta_-^2 \zeta}{4 \gamma^2 \xi}
 -  \frac{10 \mathcal{S}_+^2 \beta_-^2 \beta_+ \zeta}{3 \gamma^2 \xi}
 + \frac{5 \mathcal{S}_+^2 \beta_+^2 \zeta}{4 \gamma^2 \xi}
 -  \frac{10 \mathcal{S}_+^2 \beta_+^3 \zeta}{9 \gamma^2 \xi}
 + \frac{2 \mathcal{S}_+^2 \beta_-^2 \zeta}{9 \gamma \xi}
 + \frac{\mathcal{S}_+^2 \beta_+ \zeta}{\gamma \xi}\nonumber\\
& + \frac{2 \mathcal{S}_+^2 \beta_+^2 \zeta}{9 \gamma \xi}
 -  \frac{653 \mathcal{S}_+^2 \gamma \zeta}{504 \xi}
 -  \frac{8 \mathcal{S}_+^2 \beta_+ \gamma \zeta}{27 \xi}
 -  \frac{8 \mathcal{S}_+^2 \gamma^2 \zeta}{27 \xi}
 -  \frac{5239 \eta}{224 \xi}
 -  \frac{128 \beta_-^2 \eta}{9 \xi}
 + \frac{31 \beta_+ \eta}{9 \xi}
 -  \frac{8881 \gamma \eta}{1344 \xi}
 -  \frac{64 \beta_-^2 \gamma \eta}{9 \xi}\nonumber\\
& + \frac{31 \beta_+ \gamma \eta}{18 \xi}
 -  \frac{37 \gamma^2 \eta}{18 \xi}
 -  \frac{3425 \mathcal{S}_+^2 \zeta \eta}{4032 \xi}
 -  \frac{64 \mathcal{S}_+^2 \beta_-^2 \zeta \eta}{27 \xi}
 + \frac{31 \mathcal{S}_+^2 \beta_+ \zeta \eta}{54 \xi}
 + \frac{5 \mathcal{S}_+^2 \beta_-^2 \zeta \eta}{36 \gamma^2 \xi}
 + \frac{80 \mathcal{S}_+^2 \beta_-^2 \beta_+ \zeta \eta}{9 \gamma^2 \xi}\nonumber\\
& -  \frac{175 \mathcal{S}_+^2 \beta_+^2 \zeta \eta}{36 \gamma^2 \xi}
 + \frac{40 \mathcal{S}_+^2 \beta_+^3 \zeta \eta}{9 \gamma^2 \xi}
 + \frac{32 \mathcal{S}_+^2 \beta_-^2 \zeta \eta}{9 \gamma \xi}
 + \frac{\mathcal{S}_+^2 \beta_+ \zeta \eta}{9 \gamma \xi}
 -  \frac{40 \mathcal{S}_+^2 \beta_+^2 \zeta \eta}{9 \gamma \xi}
 -  \frac{37 \mathcal{S}_+^2 \gamma \zeta \eta}{54 \xi}
 -  \frac{295 \eta^2}{72 \xi}
 -  \frac{35 \gamma \eta^2}{144 \xi}\nonumber\\
& -  \frac{35 \mathcal{S}_+^2 \zeta \eta^2}{432 \xi}
 -  \frac{5 \mathcal{S}_+^2 \beta_+^2 \zeta \eta^2}{9 \gamma^2 \xi}
 + 4 \chi_+
 - 8 \eta \chi_+
 - 3 \beta_- \psi
 - 8 \beta_- \beta_+ \psi
 - 4 \beta_- \gamma \psi
 + \delta_- \psi
 + \beta_- \eta \psi
 -  \frac{1247 \beta_- \psi}{63 \xi^2}\nonumber\\
& -  \frac{128 \beta_- \beta_+ \psi}{9 \xi^2}
 -  \frac{565 \beta_- \gamma \psi}{21 \xi^2}
 -  \frac{128 \beta_- \beta_+ \gamma \psi}{9 \xi^2}
 -  \frac{1013 \beta_- \gamma^2 \psi}{84 \xi^2}
 -  \frac{32 \beta_- \beta_+ \gamma^2 \psi}{9 \xi^2}
 -  \frac{16 \beta_- \gamma^3 \psi}{9 \xi^2}
 -  \frac{4379 \mathcal{S}_+^2 \beta_- \zeta \psi}{1512 \xi^2}\nonumber\\
& + \frac{28 \mathcal{S}_+^2 \beta_- \beta_+ \zeta \psi}{27 \xi^2}
 + \frac{40 \mathcal{S}_+^2 \beta_-^3 \zeta \psi}{9 \gamma^2 \xi^2}
 + \frac{6235 \mathcal{S}_+^2 \beta_- \beta_+ \zeta \psi}{504 \gamma^2 \xi^2}
 + \frac{40 \mathcal{S}_+^2 \beta_- \beta_+^2 \zeta \psi}{3 \gamma^2 \xi^2}
 + \frac{1247 \mathcal{S}_+^2 \beta_- \zeta \psi}{252 \gamma \xi^2}
 + \frac{20 \mathcal{S}_+^2 \beta_-^3 \zeta \psi}{9 \gamma \xi^2}\nonumber\\
& + \frac{1987 \mathcal{S}_+^2 \beta_- \beta_+ \zeta \psi}{112 \gamma \xi^2}
 + \frac{20 \mathcal{S}_+^2 \beta_- \beta_+^2 \zeta \psi}{3 \gamma \xi^2}
 -  \frac{91 \mathcal{S}_+^2 \beta_- \gamma \zeta \psi}{18 \xi^2}
 -  \frac{64 \mathcal{S}_+^2 \beta_- \beta_+ \gamma \zeta \psi}{27 \xi^2}
 -  \frac{32 \mathcal{S}_+^2 \beta_- \gamma^2 \zeta \psi}{27 \xi^2}
 -  \frac{29 \mathcal{S}_+^4 \beta_- \zeta^2 \psi}{84 \xi^2}\nonumber\\
& -  \frac{32 \mathcal{S}_+^4 \beta_- \beta_+ \zeta^2 \psi}{81 \xi^2}
 -  \frac{25 \mathcal{S}_+^4 \beta_-^3 \beta_+ \zeta^2 \psi}{9 \gamma^4 \xi^2}
 -  \frac{25 \mathcal{S}_+^4 \beta_- \beta_+^3 \zeta^2 \psi}{9 \gamma^4 \xi^2}
 -  \frac{10 \mathcal{S}_+^4 \beta_-^3 \zeta^2 \psi}{9 \gamma^3 \xi^2}
 -  \frac{10 \mathcal{S}_+^4 \beta_- \beta_+^2 \zeta^2 \psi}{3 \gamma^3 \xi^2}
 + \frac{20 \mathcal{S}_+^4 \beta_-^3 \zeta^2 \psi}{27 \gamma^2 \xi^2}\nonumber\\
& + \frac{1529 \mathcal{S}_+^4 \beta_- \beta_+ \zeta^2 \psi}{1008 \gamma^2 \xi^2}
 + \frac{20 \mathcal{S}_+^4 \beta_- \beta_+^2 \zeta^2 \psi}{9 \gamma^2 \xi^2}
 + \frac{485 \mathcal{S}_+^4 \beta_- \zeta^2 \psi}{504 \gamma \xi^2}
 + \frac{52 \mathcal{S}_+^4 \beta_- \beta_+ \zeta^2 \psi}{27 \gamma \xi^2}
 -  \frac{16 \mathcal{S}_+^4 \beta_- \gamma \zeta^2 \psi}{81 \xi^2}
 -  \frac{140 \beta_- \eta \psi}{9 \xi^2}\nonumber\\
& -  \frac{140 \beta_- \gamma \eta \psi}{9 \xi^2}
 -  \frac{35 \beta_- \gamma^2 \eta \psi}{9 \xi^2}
 -  \frac{175 \mathcal{S}_+^2 \beta_- \zeta \eta \psi}{54 \xi^2}
 + \frac{175 \mathcal{S}_+^2 \beta_- \beta_+ \zeta \eta \psi}{18 \gamma^2 \xi^2}
 -  \frac{160 \mathcal{S}_+^2 \beta_- \beta_+^2 \zeta \eta \psi}{9 \gamma^2 \xi^2}
 + \frac{35 \mathcal{S}_+^2 \beta_- \zeta \eta \psi}{9 \gamma \xi^2}\nonumber\\
& + \frac{175 \mathcal{S}_+^2 \beta_- \beta_+ \zeta \eta \psi}{36 \gamma \xi^2}
 -  \frac{80 \mathcal{S}_+^2 \beta_- \beta_+^2 \zeta \eta \psi}{9 \gamma \xi^2}
 -  \frac{70 \mathcal{S}_+^2 \beta_- \gamma \zeta \eta \psi}{27 \xi^2}
 -  \frac{35 \mathcal{S}_+^4 \beta_- \zeta^2 \eta \psi}{81 \xi^2}
 + \frac{100 \mathcal{S}_+^4 \beta_- \beta_+^3 \zeta^2 \eta \psi}{9 \gamma^4 \xi^2}\nonumber\\
& + \frac{40 \mathcal{S}_+^4 \beta_- \beta_+^2 \zeta^2 \eta \psi}{9 \gamma^3 \xi^2}
 + \frac{175 \mathcal{S}_+^4 \beta_- \beta_+ \zeta^2 \eta \psi}{108 \gamma^2 \xi^2}
 -  \frac{80 \mathcal{S}_+^4 \beta_- \beta_+^2 \zeta^2 \eta \psi}{27 \gamma^2 \xi^2}
 + \frac{35 \mathcal{S}_+^4 \beta_- \zeta^2 \eta \psi}{54 \gamma \xi^2}
 -  \frac{239 \beta_- \psi}{252 \xi}
 -  \frac{64 \beta_- \beta_+ \psi}{9 \xi}\nonumber\\
& + \frac{73 \beta_- \gamma \psi}{56 \xi}
 -  \frac{32 \beta_- \beta_+ \gamma \psi}{9 \xi}
 + \frac{8 \beta_- \gamma^2 \psi}{9 \xi}
 -  \frac{199 \mathcal{S}_+^2 \beta_- \zeta \psi}{168 \xi}
 -  \frac{32 \mathcal{S}_+^2 \beta_- \beta_+ \zeta \psi}{27 \xi}
 + \frac{10 \mathcal{S}_+^2 \beta_-^3 \zeta \psi}{9 \gamma^2 \xi}
 -  \frac{5 \mathcal{S}_+^2 \beta_- \beta_+ \zeta \psi}{2 \gamma^2 \xi}\nonumber\\
& + \frac{10 \mathcal{S}_+^2 \beta_- \beta_+^2 \zeta \psi}{3 \gamma^2 \xi}
 -  \frac{\mathcal{S}_+^2 \beta_- \zeta \psi}{\gamma \xi}
 -  \frac{4 \mathcal{S}_+^2 \beta_- \beta_+ \zeta \psi}{9 \gamma \xi}
 + \frac{8 \mathcal{S}_+^2 \beta_- \gamma \zeta \psi}{27 \xi}
 -  \frac{31 \beta_- \eta \psi}{9 \xi}
 -  \frac{31 \beta_- \gamma \eta \psi}{18 \xi}
 -  \frac{31 \mathcal{S}_+^2 \beta_- \zeta \eta \psi}{54 \xi}\nonumber\\
& -  \frac{5 \mathcal{S}_+^2 \beta_- \beta_+ \zeta \eta \psi}{18 \gamma^2 \xi}
 -  \frac{40 \mathcal{S}_+^2 \beta_- \beta_+^2 \zeta \eta \psi}{9 \gamma^2 \xi}
 -  \frac{\mathcal{S}_+^2 \beta_- \zeta \eta \psi}{9 \gamma \xi}
 - 4 \chi_- \psi.
\end{align}
\end{subequations}

The result for the dipolar ratio $\rho_\text{dip}(x)$ is
\begin{equation}
  \rho_\text{dip} (x) = - \frac{25 \mathcal{S}_-^2 \zeta}{1536 \eta \xi^2 x^6} \left[ 1  + \rho^{\rm d}_{2} x^{} + \rho^{\rm d}_{3} x^{3/2} + \rho^{\rm d}_{4} x^{2} + \mathcal{O}(5)\right]
\end{equation}

with the coefficients
\begin{subequations}\label{eq:coeffsrhodip}
\begin{align}
\rho^{\rm d}_{2}={}&\frac{2623}{840 \xi}
 + \frac{2 \mathcal{S}_+ \beta_-}{\mathcal{S}_- \xi}
 + \frac{22 \beta_+}{3 \xi}
 + \frac{4 \mathcal{S}_+ \beta_-}{\mathcal{S}_- \gamma \xi}
 + \frac{4 \beta_+}{\gamma \xi}
 + \frac{3743 \gamma}{1680 \xi}
 + \frac{8 \beta_+ \gamma}{3 \xi}
 + \frac{\gamma^2}{3 \xi}
 + \frac{407 \mathcal{S}_+^2 \zeta}{560 \xi}
 + \frac{8 \mathcal{S}_+^2 \beta_+ \zeta}{9 \xi}
 -  \frac{5 \mathcal{S}_+^2 \beta_-^2 \zeta}{3 \gamma^2 \xi}\nonumber\\
& -  \frac{5 \mathcal{S}_+^2 \beta_+^2 \zeta}{3 \gamma^2 \xi}
 + \frac{2 \mathcal{S}_+^3 \beta_- \zeta}{3 \mathcal{S}_- \gamma \xi}
 -  \frac{2 \mathcal{S}_+^2 \beta_+ \zeta}{3 \gamma \xi}
 + \frac{\mathcal{S}_+^2 \gamma \zeta}{9 \xi}
 + \frac{13 \eta}{3 \xi}
 + \frac{13 \gamma \eta}{6 \xi}
 + \frac{13 \mathcal{S}_+^2 \zeta \eta}{18 \xi}
 + \frac{20 \mathcal{S}_+^2 \beta_+^2 \zeta \eta}{3 \gamma^2 \xi}
 -  \frac{22 \beta_- \psi}{3 \xi}\nonumber\\
& -  \frac{2 \mathcal{S}_+ \beta_+ \psi}{\mathcal{S}_- \xi}
 -  \frac{4 \beta_- \psi}{\gamma \xi}
 -  \frac{4 \mathcal{S}_+ \beta_+ \psi}{\mathcal{S}_- \gamma \xi}
 -  \frac{8 \beta_- \gamma \psi}{3 \xi}
 -  \frac{8 \mathcal{S}_+^2 \beta_- \zeta \psi}{9 \xi}
 + \frac{10 \mathcal{S}_+^2 \beta_- \beta_+ \zeta \psi}{3 \gamma^2 \xi}
 + \frac{2 \mathcal{S}_+^2 \beta_- \zeta \psi}{3 \gamma \xi}
 -  \frac{2 \mathcal{S}_+^3 \beta_+ \zeta \psi}{3 \mathcal{S}_- \gamma \xi},\\
\rho^{\rm d}_{3}={}&-2 f^{\rm ST}_{3}
 + 2 \pi
 + \pi \gamma
 -  \frac{8 \pi}{\xi},\\
\rho^{\rm d}_{4}={}&- \frac{1949}{280}
 - 2 f^{\rm ST}_{4}
 -  \frac{20 \mathcal{S}_+ \beta_-}{3 \mathcal{S}_-}
 + \frac{8}{9} \beta_-^2
 -  \frac{59}{15} \beta_+
 + \frac{8}{9} \beta_+^2
 + \frac{4 \beta_-^2}{\gamma^2}
 + \frac{16 \mathcal{S}_+ \beta_- \beta_+}{\mathcal{S}_- \gamma^2}
 + \frac{4 \beta_+^2}{\gamma^2}
 -  \frac{10 \mathcal{S}_+ \beta_-}{\mathcal{S}_- \gamma}
 + \frac{8 \beta_-^2}{3 \gamma}\nonumber\\
& -  \frac{66 \beta_+}{5 \gamma}
 + \frac{16 \mathcal{S}_+ \beta_- \beta_+}{3 \mathcal{S}_- \gamma}
 + \frac{8 \beta_+^2}{3 \gamma}
 -  \frac{133}{15} \gamma
 + \frac{44}{9} \beta_+ \gamma
 -  \frac{121}{36} \gamma^2
 + \frac{5}{3} \delta_+
 + \frac{2251}{120} \eta
 -  \frac{32}{9} \beta_-^2 \eta
 -  \frac{65}{9} \beta_+ \eta
 -  \frac{48 \beta_-^2 \eta}{\gamma^2}\nonumber\\
& -  \frac{32 \mathcal{S}_+ \beta_- \beta_+ \eta}{\mathcal{S}_- \gamma^2}
 + \frac{32 \beta_+^2 \eta}{\gamma^2}
 -  \frac{58 \mathcal{S}_+ \beta_- \eta}{3 \mathcal{S}_- \gamma}
 -  \frac{128 \beta_-^2 \eta}{3 \gamma}
 -  \frac{58 \beta_+ \eta}{3 \gamma}
 -  \frac{32 \mathcal{S}_+ \beta_- \beta_+ \eta}{3 \mathcal{S}_- \gamma}
 + \frac{32 \beta_+^2 \eta}{\gamma}
 + \frac{104}{9} \gamma \eta
 + \frac{2}{3} \gamma^2 \eta\nonumber\\
& + \frac{8}{3} \delta_+ \eta
 + \frac{55}{72} \eta^2
 + \frac{1555009}{37632 \xi^2}
 + \frac{64 \beta_-^2}{3 \xi^2}
 + \frac{1247 \beta_+}{21 \xi^2}
 + \frac{64 \beta_+^2}{3 \xi^2}
 + \frac{2672321 \gamma}{37632 \xi^2}
 + \frac{64 \beta_-^2 \gamma}{3 \xi^2}
 + \frac{565 \beta_+ \gamma}{7 \xi^2}
 + \frac{64 \beta_+^2 \gamma}{3 \xi^2}\nonumber\\
& + \frac{2275691 \gamma^2}{50176 \xi^2}
 + \frac{16 \beta_-^2 \gamma^2}{3 \xi^2}
 + \frac{1013 \beta_+ \gamma^2}{28 \xi^2}
 + \frac{16 \beta_+^2 \gamma^2}{3 \xi^2}
 + \frac{2143 \gamma^3}{168 \xi^2}
 + \frac{16 \beta_+ \gamma^3}{3 \xi^2}
 + \frac{4 \gamma^4}{3 \xi^2}
 + \frac{604795 \mathcal{S}_+^2 \zeta}{37632 \xi^2}\nonumber\\
& -  \frac{14 \mathcal{S}_+^2 \beta_-^2 \zeta}{9 \xi^2}
 + \frac{4379 \mathcal{S}_+^2 \beta_+ \zeta}{504 \xi^2}
 -  \frac{14 \mathcal{S}_+^2 \beta_+^2 \zeta}{9 \xi^2}
 -  \frac{6235 \mathcal{S}_+^2 \beta_-^2 \zeta}{336 \gamma^2 \xi^2}
 -  \frac{40 \mathcal{S}_+^2 \beta_-^2 \beta_+ \zeta}{\gamma^2 \xi^2}
 -  \frac{6235 \mathcal{S}_+^2 \beta_+^2 \zeta}{336 \gamma^2 \xi^2}
 -  \frac{40 \mathcal{S}_+^2 \beta_+^3 \zeta}{3 \gamma^2 \xi^2}\nonumber\\
& -  \frac{5961 \mathcal{S}_+^2 \beta_-^2 \zeta}{224 \gamma \xi^2}
 -  \frac{1247 \mathcal{S}_+^2 \beta_+ \zeta}{84 \gamma \xi^2}
 -  \frac{20 \mathcal{S}_+^2 \beta_-^2 \beta_+ \zeta}{\gamma \xi^2}
 -  \frac{5961 \mathcal{S}_+^2 \beta_+^2 \zeta}{224 \gamma \xi^2}
 -  \frac{20 \mathcal{S}_+^2 \beta_+^3 \zeta}{3 \gamma \xi^2}
 + \frac{4235377 \mathcal{S}_+^2 \gamma \zeta}{225792 \xi^2}\nonumber\\
& + \frac{32 \mathcal{S}_+^2 \beta_-^2 \gamma \zeta}{9 \xi^2}
 + \frac{91 \mathcal{S}_+^2 \beta_+ \gamma \zeta}{6 \xi^2}
 + \frac{32 \mathcal{S}_+^2 \beta_+^2 \gamma \zeta}{9 \xi^2}
 + \frac{257 \mathcal{S}_+^2 \gamma^2 \zeta}{36 \xi^2}
 + \frac{32 \mathcal{S}_+^2 \beta_+ \gamma^2 \zeta}{9 \xi^2}
 + \frac{8 \mathcal{S}_+^2 \gamma^3 \zeta}{9 \xi^2}
 + \frac{235225 \mathcal{S}_+^4 \zeta^2}{150528 \xi^2}\nonumber\\
& + \frac{16 \mathcal{S}_+^4 \beta_-^2 \zeta^2}{27 \xi^2}
 + \frac{29 \mathcal{S}_+^4 \beta_+ \zeta^2}{28 \xi^2}
 + \frac{16 \mathcal{S}_+^4 \beta_+^2 \zeta^2}{27 \xi^2}
 + \frac{25 \mathcal{S}_+^4 \beta_-^4 \zeta^2}{12 \gamma^4 \xi^2}
 + \frac{25 \mathcal{S}_+^4 \beta_-^2 \beta_+^2 \zeta^2}{2 \gamma^4 \xi^2}
 + \frac{25 \mathcal{S}_+^4 \beta_+^4 \zeta^2}{12 \gamma^4 \xi^2}
 + \frac{10 \mathcal{S}_+^4 \beta_-^2 \beta_+ \zeta^2}{\gamma^3 \xi^2}\nonumber\\
& + \frac{10 \mathcal{S}_+^4 \beta_+^3 \zeta^2}{3 \gamma^3 \xi^2}
 -  \frac{1529 \mathcal{S}_+^4 \beta_-^2 \zeta^2}{672 \gamma^2 \xi^2}
 -  \frac{20 \mathcal{S}_+^4 \beta_-^2 \beta_+ \zeta^2}{3 \gamma^2 \xi^2}
 -  \frac{1529 \mathcal{S}_+^4 \beta_+^2 \zeta^2}{672 \gamma^2 \xi^2}
 -  \frac{20 \mathcal{S}_+^4 \beta_+^3 \zeta^2}{9 \gamma^2 \xi^2}
 -  \frac{26 \mathcal{S}_+^4 \beta_-^2 \zeta^2}{9 \gamma \xi^2}\nonumber\\
& -  \frac{485 \mathcal{S}_+^4 \beta_+ \zeta^2}{168 \gamma \xi^2}
 -  \frac{26 \mathcal{S}_+^4 \beta_+^2 \zeta^2}{9 \gamma \xi^2}
 + \frac{485 \mathcal{S}_+^4 \gamma \zeta^2}{504 \xi^2}
 + \frac{16 \mathcal{S}_+^4 \beta_+ \gamma \zeta^2}{27 \xi^2}
 + \frac{4 \mathcal{S}_+^4 \gamma^2 \zeta^2}{27 \xi^2}
 + \frac{6235 \eta}{96 \xi^2}
 -  \frac{256 \beta_-^2 \eta}{3 \xi^2}
 + \frac{140 \beta_+ \eta}{3 \xi^2}\nonumber\\
& + \frac{2825 \gamma \eta}{32 \xi^2}
 -  \frac{256 \beta_-^2 \gamma \eta}{3 \xi^2}
 + \frac{140 \beta_+ \gamma \eta}{3 \xi^2}
 + \frac{5065 \gamma^2 \eta}{128 \xi^2}
 -  \frac{64 \beta_-^2 \gamma^2 \eta}{3 \xi^2}
 + \frac{35 \beta_+ \gamma^2 \eta}{3 \xi^2}
 + \frac{35 \gamma^3 \eta}{6 \xi^2}
 + \frac{6755 \mathcal{S}_+^2 \zeta \eta}{288 \xi^2}\nonumber\\
& -  \frac{64 \mathcal{S}_+^2 \beta_-^2 \zeta \eta}{9 \xi^2}
 + \frac{175 \mathcal{S}_+^2 \beta_+ \zeta \eta}{18 \xi^2}
 + \frac{40 \mathcal{S}_+^2 \beta_+^2 \zeta \eta}{3 \xi^2}
 -  \frac{175 \mathcal{S}_+^2 \beta_-^2 \zeta \eta}{12 \gamma^2 \xi^2}
 + \frac{320 \mathcal{S}_+^2 \beta_-^2 \beta_+ \zeta \eta}{3 \gamma^2 \xi^2}
 + \frac{835 \mathcal{S}_+^2 \beta_+^2 \zeta \eta}{14 \gamma^2 \xi^2}\nonumber\\
& + \frac{160 \mathcal{S}_+^2 \beta_+^3 \zeta \eta}{3 \gamma^2 \xi^2}
 + \frac{283 \mathcal{S}_+^2 \beta_-^2 \zeta \eta}{8 \gamma \xi^2}
 -  \frac{35 \mathcal{S}_+^2 \beta_+ \zeta \eta}{3 \gamma \xi^2}
 + \frac{160 \mathcal{S}_+^2 \beta_-^2 \beta_+ \zeta \eta}{3 \gamma \xi^2}
 + \frac{4745 \mathcal{S}_+^2 \beta_+^2 \zeta \eta}{84 \gamma \xi^2}
 + \frac{80 \mathcal{S}_+^2 \beta_+^3 \zeta \eta}{3 \gamma \xi^2}\nonumber\\
& + \frac{3745 \mathcal{S}_+^2 \gamma \zeta \eta}{192 \xi^2}
 -  \frac{128 \mathcal{S}_+^2 \beta_-^2 \gamma \zeta \eta}{9 \xi^2}
 + \frac{70 \mathcal{S}_+^2 \beta_+ \gamma \zeta \eta}{9 \xi^2}
 + \frac{35 \mathcal{S}_+^2 \gamma^2 \zeta \eta}{9 \xi^2}
 + \frac{2425 \mathcal{S}_+^4 \zeta^2 \eta}{1152 \xi^2}
 -  \frac{64 \mathcal{S}_+^4 \beta_-^2 \zeta^2 \eta}{27 \xi^2}\nonumber\\
& + \frac{35 \mathcal{S}_+^4 \beta_+ \zeta^2 \eta}{27 \xi^2}
 -  \frac{50 \mathcal{S}_+^4 \beta_-^2 \beta_+^2 \zeta^2 \eta}{\gamma^4 \xi^2}
 -  \frac{50 \mathcal{S}_+^4 \beta_+^4 \zeta^2 \eta}{3 \gamma^4 \xi^2}
 -  \frac{80 \mathcal{S}_+^4 \beta_-^2 \beta_+ \zeta^2 \eta}{3 \gamma^3 \xi^2}
 -  \frac{40 \mathcal{S}_+^4 \beta_+^3 \zeta^2 \eta}{3 \gamma^3 \xi^2}
 -  \frac{559 \mathcal{S}_+^4 \beta_-^2 \zeta^2 \eta}{72 \gamma^2 \xi^2}\nonumber\\
& + \frac{160 \mathcal{S}_+^4 \beta_-^2 \beta_+ \zeta^2 \eta}{9 \gamma^2 \xi^2}
 + \frac{3025 \mathcal{S}_+^4 \beta_+^2 \zeta^2 \eta}{252 \gamma^2 \xi^2}
 + \frac{80 \mathcal{S}_+^4 \beta_+^3 \zeta^2 \eta}{9 \gamma^2 \xi^2}
 + \frac{64 \mathcal{S}_+^4 \beta_-^2 \zeta^2 \eta}{9 \gamma \xi^2}
 -  \frac{35 \mathcal{S}_+^4 \beta_+ \zeta^2 \eta}{18 \gamma \xi^2}
 + \frac{40 \mathcal{S}_+^4 \beta_+^2 \zeta^2 \eta}{9 \gamma \xi^2}\nonumber\\
& + \frac{35 \mathcal{S}_+^4 \gamma \zeta^2 \eta}{54 \xi^2}
 + \frac{1225 \eta^2}{48 \xi^2}
 + \frac{1225 \gamma \eta^2}{48 \xi^2}
 + \frac{1225 \gamma^2 \eta^2}{192 \xi^2}
 + \frac{1225 \mathcal{S}_+^2 \zeta \eta^2}{144 \xi^2}
 + \frac{175 \mathcal{S}_+^2 \beta_+^2 \zeta \eta^2}{3 \gamma^2 \xi^2}
 + \frac{175 \mathcal{S}_+^2 \beta_+^2 \zeta \eta^2}{6 \gamma \xi^2}\nonumber\\
& + \frac{1225 \mathcal{S}_+^2 \gamma \zeta \eta^2}{288 \xi^2}
 + \frac{1225 \mathcal{S}_+^4 \zeta^2 \eta^2}{1728 \xi^2}
 + \frac{100 \mathcal{S}_+^4 \beta_+^4 \zeta^2 \eta^2}{3 \gamma^4 \xi^2}
 + \frac{175 \mathcal{S}_+^4 \beta_+^2 \zeta^2 \eta^2}{18 \gamma^2 \xi^2}
 -  \frac{142951}{6480 \xi}
 + \frac{2143 \mathcal{S}_+ \beta_-}{84 \mathcal{S}_- \xi}
 + \frac{32 \beta_-^2}{3 \xi}\nonumber\\
& + \frac{361 \beta_+}{140 \xi}
 + \frac{64 \mathcal{S}_+ \beta_- \beta_+}{3 \mathcal{S}_- \xi}
 + \frac{32 \beta_+^2}{3 \xi}
 + \frac{1247 \mathcal{S}_+ \beta_-}{42 \mathcal{S}_- \gamma \xi}
 + \frac{64 \beta_-^2}{3 \gamma \xi}
 + \frac{1247 \beta_+}{42 \gamma \xi}
 + \frac{128 \mathcal{S}_+ \beta_- \beta_+}{3 \mathcal{S}_- \gamma \xi}
 + \frac{64 \beta_+^2}{3 \gamma \xi}
 -  \frac{47343 \gamma}{1120 \xi}\nonumber\\
& + \frac{16 \mathcal{S}_+ \beta_- \gamma}{3 \mathcal{S}_- \xi}
 -  \frac{84 \beta_+ \gamma}{5 \xi}
 -  \frac{5177 \gamma^2}{280 \xi}
 -  \frac{16 \beta_+ \gamma^2}{3 \xi}
 -  \frac{8 \gamma^3}{3 \xi}
 -  \frac{4171 \mathcal{S}_+^2 \zeta}{672 \xi}
 + \frac{16 \mathcal{S}_+^3 \beta_- \zeta}{9 \mathcal{S}_- \xi}
 + \frac{28 \mathcal{S}_+^2 \beta_+ \zeta}{45 \xi}
 -  \frac{20 \mathcal{S}_+^3 \beta_-^3 \zeta}{3 \mathcal{S}_- \gamma^3 \xi}\nonumber\\
& -  \frac{20 \mathcal{S}_+^2 \beta_-^2 \beta_+ \zeta}{\gamma^3 \xi}
 -  \frac{20 \mathcal{S}_+^3 \beta_- \beta_+^2 \zeta}{\mathcal{S}_- \gamma^3 \xi}
 -  \frac{20 \mathcal{S}_+^2 \beta_+^3 \zeta}{3 \gamma^3 \xi}
 + \frac{11 \mathcal{S}_+^2 \beta_-^2 \zeta}{6 \gamma^2 \xi}
 -  \frac{32 \mathcal{S}_+^3 \beta_- \beta_+ \zeta}{3 \mathcal{S}_- \gamma^2 \xi}
 + \frac{11 \mathcal{S}_+^2 \beta_+^2 \zeta}{6 \gamma^2 \xi}
 + \frac{485 \mathcal{S}_+^3 \beta_- \zeta}{84 \mathcal{S}_- \gamma \xi}\nonumber\\
& + \frac{62 \mathcal{S}_+^2 \beta_-^2 \zeta}{9 \gamma \xi}
 + \frac{1611 \mathcal{S}_+^2 \beta_+ \zeta}{140 \gamma \xi}
 + \frac{64 \mathcal{S}_+^3 \beta_- \beta_+ \zeta}{9 \mathcal{S}_- \gamma \xi}
 + \frac{62 \mathcal{S}_+^2 \beta_+^2 \zeta}{9 \gamma \xi}
 -  \frac{12091 \mathcal{S}_+^2 \gamma \zeta}{2520 \xi}
 -  \frac{16 \mathcal{S}_+^2 \beta_+ \gamma \zeta}{9 \xi}
 -  \frac{8 \mathcal{S}_+^2 \gamma^2 \zeta}{9 \xi}\nonumber\\
& -  \frac{10513 \eta}{144 \xi}
 + \frac{35 \mathcal{S}_+ \beta_- \eta}{3 \mathcal{S}_- \xi}
 -  \frac{128 \beta_-^2 \eta}{3 \xi}
 + \frac{11 \beta_+ \eta}{3 \xi}
 -  \frac{128 \mathcal{S}_+ \beta_- \beta_+ \eta}{3 \mathcal{S}_- \xi}
 + \frac{70 \mathcal{S}_+ \beta_- \eta}{3 \mathcal{S}_- \gamma \xi}
 -  \frac{256 \beta_-^2 \eta}{3 \gamma \xi}
 + \frac{70 \beta_+ \eta}{3 \gamma \xi}\nonumber\\
& -  \frac{256 \mathcal{S}_+ \beta_- \beta_+ \eta}{3 \mathcal{S}_- \gamma \xi}
 -  \frac{22697 \gamma \eta}{672 \xi}
 -  \frac{4 \beta_+ \gamma \eta}{\xi}
 -  \frac{47 \gamma^2 \eta}{6 \xi}
 -  \frac{12793 \mathcal{S}_+^2 \zeta \eta}{2016 \xi}
 -  \frac{4 \mathcal{S}_+^2 \beta_+ \zeta \eta}{3 \xi}
 + \frac{160 \mathcal{S}_+^2 \beta_-^2 \beta_+ \zeta \eta}{3 \gamma^3 \xi}\nonumber\\
& + \frac{80 \mathcal{S}_+^3 \beta_- \beta_+^2 \zeta \eta}{\mathcal{S}_- \gamma^3 \xi}
 + \frac{80 \mathcal{S}_+^2 \beta_+^3 \zeta \eta}{3 \gamma^3 \xi}
 + \frac{143 \mathcal{S}_+^2 \beta_-^2 \zeta \eta}{6 \gamma^2 \xi}
 + \frac{64 \mathcal{S}_+^3 \beta_- \beta_+ \zeta \eta}{3 \mathcal{S}_- \gamma^2 \xi}
 -  \frac{157 \mathcal{S}_+^2 \beta_+^2 \zeta \eta}{6 \gamma^2 \xi}
 + \frac{35 \mathcal{S}_+^3 \beta_- \zeta \eta}{9 \mathcal{S}_- \gamma \xi}\nonumber\\
& -  \frac{128 \mathcal{S}_+^2 \beta_-^2 \zeta \eta}{9 \gamma \xi}
 + \frac{53 \mathcal{S}_+^2 \beta_+ \zeta \eta}{9 \gamma \xi}
 -  \frac{128 \mathcal{S}_+^3 \beta_- \beta_+ \zeta \eta}{9 \mathcal{S}_- \gamma \xi}
 -  \frac{40 \mathcal{S}_+^2 \beta_+^2 \zeta \eta}{3 \gamma \xi}
 -  \frac{47 \mathcal{S}_+^2 \gamma \zeta \eta}{18 \xi}
 -  \frac{575 \eta^2}{36 \xi}
 -  \frac{35 \gamma \eta^2}{8 \xi}\nonumber\\
& -  \frac{35 \mathcal{S}_+^2 \zeta \eta^2}{24 \xi}
 -  \frac{10 \mathcal{S}_+^2 \beta_+^2 \zeta \eta^2}{\gamma^2 \xi}
 + \frac{4 \mathcal{S}_+ \chi_-}{\mathcal{S}_- \gamma}
 -  \frac{8 \mathcal{S}_+ \eta \chi_-}{\mathcal{S}_- \gamma}
 + \frac{8}{3} \chi_+
 + \frac{4 \chi_+}{\gamma}
 -  \frac{16}{3} \eta \chi_+
 -  \frac{8 \eta \chi_+}{\gamma}
 + \frac{11 \mathcal{S}_+ \psi}{2 \mathcal{S}_-}
 + \frac{59}{15} \beta_- \psi\nonumber\\
& + \frac{20 \mathcal{S}_+ \beta_+ \psi}{3 \mathcal{S}_-}
 -  \frac{16}{9} \beta_- \beta_+ \psi
 -  \frac{8 \mathcal{S}_+ \beta_-^2 \psi}{\mathcal{S}_- \gamma^2}
 -  \frac{8 \beta_- \beta_+ \psi}{\gamma^2}
 -  \frac{8 \mathcal{S}_+ \beta_+^2 \psi}{\mathcal{S}_- \gamma^2}
 + \frac{66 \beta_- \psi}{5 \gamma}
 -  \frac{8 \mathcal{S}_+ \beta_-^2 \psi}{3 \mathcal{S}_- \gamma}
 + \frac{10 \mathcal{S}_+ \beta_+ \psi}{\mathcal{S}_- \gamma}\nonumber\\
& -  \frac{16 \beta_- \beta_+ \psi}{3 \gamma}
 -  \frac{8 \mathcal{S}_+ \beta_+^2 \psi}{3 \mathcal{S}_- \gamma}
 + \frac{3 \mathcal{S}_+ \gamma \psi}{\mathcal{S}_-}
 -  \frac{44}{9} \beta_- \gamma \psi
 + \frac{5}{3} \delta_- \psi
 -  \frac{25}{9} \beta_- \eta \psi
 + \frac{22 \beta_- \eta \psi}{3 \gamma}
 + \frac{22 \mathcal{S}_+ \beta_+ \eta \psi}{3 \mathcal{S}_- \gamma}\nonumber\\
& -  \frac{1247 \beta_- \psi}{21 \xi^2}
 -  \frac{128 \beta_- \beta_+ \psi}{3 \xi^2}
 -  \frac{565 \beta_- \gamma \psi}{7 \xi^2}
 -  \frac{128 \beta_- \beta_+ \gamma \psi}{3 \xi^2}
 -  \frac{1013 \beta_- \gamma^2 \psi}{28 \xi^2}
 -  \frac{32 \beta_- \beta_+ \gamma^2 \psi}{3 \xi^2}
 -  \frac{16 \beta_- \gamma^3 \psi}{3 \xi^2}\nonumber\\
& -  \frac{4379 \mathcal{S}_+^2 \beta_- \zeta \psi}{504 \xi^2}
 + \frac{28 \mathcal{S}_+^2 \beta_- \beta_+ \zeta \psi}{9 \xi^2}
 + \frac{40 \mathcal{S}_+^2 \beta_-^3 \zeta \psi}{3 \gamma^2 \xi^2}
 + \frac{6235 \mathcal{S}_+^2 \beta_- \beta_+ \zeta \psi}{168 \gamma^2 \xi^2}
 + \frac{40 \mathcal{S}_+^2 \beta_- \beta_+^2 \zeta \psi}{\gamma^2 \xi^2}
 + \frac{1247 \mathcal{S}_+^2 \beta_- \zeta \psi}{84 \gamma \xi^2}\nonumber\\
& + \frac{20 \mathcal{S}_+^2 \beta_-^3 \zeta \psi}{3 \gamma \xi^2}
 + \frac{5961 \mathcal{S}_+^2 \beta_- \beta_+ \zeta \psi}{112 \gamma \xi^2}
 + \frac{20 \mathcal{S}_+^2 \beta_- \beta_+^2 \zeta \psi}{\gamma \xi^2}
 -  \frac{91 \mathcal{S}_+^2 \beta_- \gamma \zeta \psi}{6 \xi^2}
 -  \frac{64 \mathcal{S}_+^2 \beta_- \beta_+ \gamma \zeta \psi}{9 \xi^2}
 -  \frac{32 \mathcal{S}_+^2 \beta_- \gamma^2 \zeta \psi}{9 \xi^2}\nonumber\\
& -  \frac{29 \mathcal{S}_+^4 \beta_- \zeta^2 \psi}{28 \xi^2}
 -  \frac{32 \mathcal{S}_+^4 \beta_- \beta_+ \zeta^2 \psi}{27 \xi^2}
 -  \frac{25 \mathcal{S}_+^4 \beta_-^3 \beta_+ \zeta^2 \psi}{3 \gamma^4 \xi^2}
 -  \frac{25 \mathcal{S}_+^4 \beta_- \beta_+^3 \zeta^2 \psi}{3 \gamma^4 \xi^2}
 -  \frac{10 \mathcal{S}_+^4 \beta_-^3 \zeta^2 \psi}{3 \gamma^3 \xi^2}
 -  \frac{10 \mathcal{S}_+^4 \beta_- \beta_+^2 \zeta^2 \psi}{\gamma^3 \xi^2}\nonumber\\
& + \frac{20 \mathcal{S}_+^4 \beta_-^3 \zeta^2 \psi}{9 \gamma^2 \xi^2}
 + \frac{1529 \mathcal{S}_+^4 \beta_- \beta_+ \zeta^2 \psi}{336 \gamma^2 \xi^2}
 + \frac{20 \mathcal{S}_+^4 \beta_- \beta_+^2 \zeta^2 \psi}{3 \gamma^2 \xi^2}
 + \frac{485 \mathcal{S}_+^4 \beta_- \zeta^2 \psi}{168 \gamma \xi^2}
 + \frac{52 \mathcal{S}_+^4 \beta_- \beta_+ \zeta^2 \psi}{9 \gamma \xi^2}\nonumber\\
& -  \frac{16 \mathcal{S}_+^4 \beta_- \gamma \zeta^2 \psi}{27 \xi^2}
 -  \frac{140 \beta_- \eta \psi}{3 \xi^2}
 -  \frac{140 \beta_- \gamma \eta \psi}{3 \xi^2}
 -  \frac{35 \beta_- \gamma^2 \eta \psi}{3 \xi^2}
 -  \frac{175 \mathcal{S}_+^2 \beta_- \zeta \eta \psi}{18 \xi^2}
 + \frac{175 \mathcal{S}_+^2 \beta_- \beta_+ \zeta \eta \psi}{6 \gamma^2 \xi^2}\nonumber\\
& -  \frac{160 \mathcal{S}_+^2 \beta_- \beta_+^2 \zeta \eta \psi}{3 \gamma^2 \xi^2}
 + \frac{35 \mathcal{S}_+^2 \beta_- \zeta \eta \psi}{3 \gamma \xi^2}
 + \frac{175 \mathcal{S}_+^2 \beta_- \beta_+ \zeta \eta \psi}{12 \gamma \xi^2}
 -  \frac{80 \mathcal{S}_+^2 \beta_- \beta_+^2 \zeta \eta \psi}{3 \gamma \xi^2}
 -  \frac{70 \mathcal{S}_+^2 \beta_- \gamma \zeta \eta \psi}{9 \xi^2}\nonumber\\
& -  \frac{35 \mathcal{S}_+^4 \beta_- \zeta^2 \eta \psi}{27 \xi^2}
 + \frac{100 \mathcal{S}_+^4 \beta_- \beta_+^3 \zeta^2 \eta \psi}{3 \gamma^4 \xi^2}
 + \frac{40 \mathcal{S}_+^4 \beta_- \beta_+^2 \zeta^2 \eta \psi}{3 \gamma^3 \xi^2}
 + \frac{175 \mathcal{S}_+^4 \beta_- \beta_+ \zeta^2 \eta \psi}{36 \gamma^2 \xi^2}
 -  \frac{80 \mathcal{S}_+^4 \beta_- \beta_+^2 \zeta^2 \eta \psi}{9 \gamma^2 \xi^2}\nonumber\\
& + \frac{35 \mathcal{S}_+^4 \beta_- \zeta^2 \eta \psi}{18 \gamma \xi^2}
 -  \frac{361 \beta_- \psi}{140 \xi}
 -  \frac{32 \mathcal{S}_+ \beta_-^2 \psi}{3 \mathcal{S}_- \xi}
 -  \frac{2143 \mathcal{S}_+ \beta_+ \psi}{84 \mathcal{S}_- \xi}
 -  \frac{64 \beta_- \beta_+ \psi}{3 \xi}
 -  \frac{32 \mathcal{S}_+ \beta_+^2 \psi}{3 \mathcal{S}_- \xi}
 -  \frac{1247 \beta_- \psi}{42 \gamma \xi}\nonumber\\
& -  \frac{64 \mathcal{S}_+ \beta_-^2 \psi}{3 \mathcal{S}_- \gamma \xi}
 -  \frac{1247 \mathcal{S}_+ \beta_+ \psi}{42 \mathcal{S}_- \gamma \xi}
 -  \frac{128 \beta_- \beta_+ \psi}{3 \gamma \xi}
 -  \frac{64 \mathcal{S}_+ \beta_+^2 \psi}{3 \mathcal{S}_- \gamma \xi}
 + \frac{84 \beta_- \gamma \psi}{5 \xi}
 -  \frac{16 \mathcal{S}_+ \beta_+ \gamma \psi}{3 \mathcal{S}_- \xi}
 + \frac{16 \beta_- \gamma^2 \psi}{3 \xi}\nonumber\\
& -  \frac{28 \mathcal{S}_+^2 \beta_- \zeta \psi}{45 \xi}
 -  \frac{16 \mathcal{S}_+^3 \beta_+ \zeta \psi}{9 \mathcal{S}_- \xi}
 + \frac{20 \mathcal{S}_+^2 \beta_-^3 \zeta \psi}{3 \gamma^3 \xi}
 + \frac{20 \mathcal{S}_+^3 \beta_-^2 \beta_+ \zeta \psi}{\mathcal{S}_- \gamma^3 \xi}
 + \frac{20 \mathcal{S}_+^2 \beta_- \beta_+^2 \zeta \psi}{\gamma^3 \xi}
 + \frac{20 \mathcal{S}_+^3 \beta_+^3 \zeta \psi}{3 \mathcal{S}_- \gamma^3 \xi}\nonumber\\
& + \frac{16 \mathcal{S}_+^3 \beta_-^2 \zeta \psi}{3 \mathcal{S}_- \gamma^2 \xi}
 -  \frac{11 \mathcal{S}_+^2 \beta_- \beta_+ \zeta \psi}{3 \gamma^2 \xi}
 + \frac{16 \mathcal{S}_+^3 \beta_+^2 \zeta \psi}{3 \mathcal{S}_- \gamma^2 \xi}
 -  \frac{1611 \mathcal{S}_+^2 \beta_- \zeta \psi}{140 \gamma \xi}
 -  \frac{32 \mathcal{S}_+^3 \beta_-^2 \zeta \psi}{9 \mathcal{S}_- \gamma \xi}
 -  \frac{485 \mathcal{S}_+^3 \beta_+ \zeta \psi}{84 \mathcal{S}_- \gamma \xi}\nonumber\\
& -  \frac{124 \mathcal{S}_+^2 \beta_- \beta_+ \zeta \psi}{9 \gamma \xi}
 -  \frac{32 \mathcal{S}_+^3 \beta_+^2 \zeta \psi}{9 \mathcal{S}_- \gamma \xi}
 + \frac{16 \mathcal{S}_+^2 \beta_- \gamma \zeta \psi}{9 \xi}
 -  \frac{11 \beta_- \eta \psi}{3 \xi}
 -  \frac{35 \mathcal{S}_+ \beta_+ \eta \psi}{3 \mathcal{S}_- \xi}
 -  \frac{70 \beta_- \eta \psi}{3 \gamma \xi}
 -  \frac{70 \mathcal{S}_+ \beta_+ \eta \psi}{3 \mathcal{S}_- \gamma \xi}\nonumber\\
& + \frac{4 \beta_- \gamma \eta \psi}{\xi}
 + \frac{4 \mathcal{S}_+^2 \beta_- \zeta \eta \psi}{3 \xi}
 -  \frac{80 \mathcal{S}_+^2 \beta_- \beta_+^2 \zeta \eta \psi}{3 \gamma^3 \xi}
 -  \frac{80 \mathcal{S}_+^3 \beta_+^3 \zeta \eta \psi}{3 \mathcal{S}_- \gamma^3 \xi}
 -  \frac{5 \mathcal{S}_+^2 \beta_- \beta_+ \zeta \eta \psi}{\gamma^2 \xi}
 -  \frac{53 \mathcal{S}_+^2 \beta_- \zeta \eta \psi}{9 \gamma \xi}\nonumber\\
& -  \frac{35 \mathcal{S}_+^3 \beta_+ \zeta \eta \psi}{9 \mathcal{S}_- \gamma \xi}
 -  \frac{8}{3} \chi_- \psi
 -  \frac{4 \chi_- \psi}{\gamma}
 -  \frac{4 \mathcal{S}_+ \chi_+ \psi}{\mathcal{S}_- \gamma}.
\end{align}
\end{subequations}

\end{widetext}

%%%%%%%%%%%%%%%%%%%%%%%%%%%%%%%%%%%%

\bibliography{reference,ST_waveform}
\end{document}